\documentclass[11pt,letterpaper]{article}
\usepackage[letterpaper, margin=0.75in]{geometry}
\usepackage{graphicx} 
\usepackage{amsmath}

\usepackage{xcolor}

\renewcommand{\epsilon}{\varepsilon}

\renewcommand{\d}{\mathrm{d}}

\renewcommand{\epsilon}{\varepsilon}

\newcommand{\TimeDeriv}{\frac{\textrm{d}}{\textrm{dt}}}

\allowdisplaybreaks

\usepackage[square,numbers,sort&compress]{natbib}
\usepackage{comment}
\usepackage{graphicx}
\usepackage{overpic} 
\usepackage[normalem]{ulem}

 \graphicspath{ {./Figures/} }
 
 \usepackage{setspace}
 \usepackage[right,mathlines]{lineno}
\let\oldequation\equation
\let\oldendequation\endequation

\renewenvironment{equation}
  {\linenomathNonumbers\oldequation}
  {\oldendequation\endlinenomath}
  
\let\oldalign\align
\let\oldendalign\endalign

\renewenvironment{align}
  {\linenomathNonumbers\oldalign}
  {\oldendalign\endlinenomath}

\renewenvironment{align*}
  {\linenomathNonumbers\oldalign\notag}
  {\notag \oldendalign \endlinenomath}   

    \graphicspath{ {./Figures/} }
 
\title{A nonparametric approach to practical identifiability of nonlinear mixed effects models}
\author{Tyler Cassidy\textsuperscript{1,*}, Stuart T. Johnston\textsuperscript{2}, Michael Plank\textsuperscript{3}, Imke Botha\textsuperscript{2},  \\ 
Jennifer A. Flegg\textsuperscript{2}, Ryan J. Murphy\textsuperscript{4}, Sara Hamis\textsuperscript{5}  
     }
\date{\today}

\begin{document}
\maketitle
\small{ 
\noindent \textsuperscript{1} University of Leeds, Leeds,  United Kingdom. 
\textsuperscript{2} University of Melbourne, Melbourne, Australia 
\textsuperscript{3} University of Canterbury, Christchurch, New Zealand 
\textsuperscript{4}UniSA STEM, The University of South Australia, Mawson Lakes, South Australia 5095, Australia 
\textsuperscript{5} Uppsala University, Uppsala, Sweden \\ 
Correspondence to: t.cassidy1@leeds.ac.uk
}

\section*{Abstract}
Mathematical modelling is a widely used approach to understand and interpret clinical trial data. This modelling typically involves fitting mechanistic mathematical models to data from individual trial participants. Despite the widespread adoption of this individual-based fitting, it is becoming increasingly common to take a hierarchical approach to parameter estimation, where modellers characterize the population parameter distributions, rather than considering each individual independently. This hierarchical parameter estimation is standard in pharmacometric modelling. However, many of the existing techniques for parameter identifiability do not immediately translate from the individual-based fitting to the hierarchical setting. In this work, we propose a nonparametric approach to study practical identifiability within a hierarchical parameter estimation framework. We focus on the commonly used nonlinear mixed effects framework and investigate two well-studied examples from the pharmacometrics and viral dynamics literature to illustrate the potential utility of our approach.  

\section{Introduction} 

Mathematical modelling has become an important approach to understand and interpret clinical trial data \citep{Sher2022,Perelson2002,Hill2018,Smith2018}. Typical mathematical models involve a number of unknown parameters, each representing a distinct biological mechanism, which are estimated by fitting the model to clinical trial data \citep{Goyal2019,Mackey2020,Vemparala2024a,Craig2023}. When considering the dynamics of distinct participants, this parameter estimation involves determining a set of model parameters that ensures that the model best represents the dynamics of each individual participant \citep{Cassidy2023,Cardozo-Ojeda2021,Perelson1996,Zhang2022}. Consequently, individual trial participants are represented by a corresponding set of model parameters. However, for nonlinear mechanistic models commonly used to interpret clinical data \citep{Braniff2024,Reeves2023,Jenner2021,Jenner2020,Goyal2019}, it is possible that the model parameters are not uniquely determined by the available data for each participant. Consequently, there has been increased interest throughout the field of mathematical biology to understand the identifiability of mechanistic models \citep{Frohlich2023,Aunins2018,Prague2019,Iyaniwura2024}.

Broadly speaking, identifiability analysis attempts to determines if the mapping from model parameterization to observable data is one-to-one. If this mapping is one-to-one in the idealized case where we observe every model output continuously, then the model is \textit{structurally} identifiable. Structural identifiability is a property of the model structure and is independent of the experimental data available for parameterization \citep{Craig2023,Raue2009}. Conversely, a mathematical model is \textit{practically} identifiable if the model can be uniquely parameterized from available experimental data \citep{Kreutz2009,Raue2009,Simpson2024,Miao2011}. However, directly showing that available clinical data is sufficient to uniquely parameterize a mathematical model is difficult. Consequently, several analytic and computational approaches have been developed to study the identifiability of a mathematical model \citep{Dong2023,Raue2009,Maiwald2016,Sharp2022,Ciocanel2024,Bellu2007,Raue2014,Stapor2018a}. In general, these existing tools focus on determining if the observed individual-level dynamics are sufficient to constrain each individual parameterization.
 
However, hierarchical parameter estimation approaches are becoming increasingly common in the mathematical modelling of clinical trial data. Rather than estimating a parameterisation for each individual participant, hierarchical approaches focus on determining  multi-level parameter distributions. For example, nonlinear mixed effects (NLME) models are extensively used throughout the pharmacokinetics community, and the NLME framework is increasingly used in the development of mathematical models to understand clinical trial data \citep{Iyaniwura2024,CortesRios2025,Marc2023,Traynard2020,Lavielle2014}. These NLME models are designed to capture the inter-individual variability within a clinically relevant population \citep{Lavielle2014,Kuhn2005} by simultaneously considering the totality of the individual-level dynamics to estimate the population-level parameter distributions. Essentially, the NLME framework leverages the entirety of the individual clinical data to constrain the population-level distributions. These population-level distributions then act as priors when determining the individual parameter sets for each participant \citep{Kuhn2005,Guedj2011,Lavielle2014,Craig2023}. Consequently, each individual participant represents a sample from the underlying population-level distribution.


Many recent advances in parameter identifiability have focused on individual-level dynamics and do not immediately extend to hierarchical modelling \citep{Raue2009,Raue2014}. Consequently, there has been recent interest in extending existing identifiability techniques to a hierarchical modelling framework \citep{Shivva2013,Janzen2016,Janzen2018}.  For example, \citet{Lavielle2016} illustrated how models which are unidentifiable on the individual level may become identifiable at the population level, with examples drawn from pharmacokinetic modelling. In similar work, \citet{Duchesne2021} leveraged the parametric representation of population parameter distributions to facilitate model reduction in an \textit{in vitro} model of erythropoiesis, while \citet{ElMessaoudi2023} used a simulation-based approach to re-estimate the known parametric population distributions from simulated data. In each of these cases, the identifiability of the NLME model was assessed via estimates of the parametric representation of the population parameter distribution (for example, estimating the mean and variance of a lognormal distribution). This parametric perspective on identifiability of NLME models is a natural extension of the familiar definition of identifiability for non-hierarchical models \citep{Janzen2016,Lavielle2016}. However, this approach relies on \textit{a priori} fixing a parametric representation of the population distribution and implicitly assumes that the individual participants can precisely determine the underlying patient population. Here, we propose a nonparametric approach to perform practical identifiability analysis of NLME models.

Rather than determining if the individual samples precisely determine the population distribution, we wish to quantify if we can distinguish between the estimated population distributions. Informally, for two population parameter distributions that equally well describe the data, rather than asking if these two distributions are precisely equal, we instead wish to determine when they are different in some statistically significant way. We attempt to distinguish between these two distributions from the individual and population perspective. At the individual level, we use the Kolmogorov-Smirnov two sample test to determine if the individual estimates distinguish between the population parameter distributions. This approach identifies if the individual-level data is sufficient to distinguish between the underlying population parameter distributions. At the population level, we use the overlapping index of the two distributions \citep{Schmid2006,Pastore2019}, which is related to the total variation difference between two probability measures, and measures the statistical difference between the distributions. We illustrate the proposed approach using a common example from pharmacology, the Friberg model of chemotherapy-induced neutropenia \citep{Friberg2002}, and a well-known example from mathematical biology, the standard viral dynamics model \citep{Perelson2002}. 

Both the Friberg and standard viral dynamics models have been extensively well-studied \citep{Wu2008,Miao2011,Evans2018} and the sufficient conditions for the identifiability of both these models, when considering single-subject fits, are known. However, as shown by \citet{Lavielle2014}, the inter-individual variability present in clinical data may result in a model that is identifiable at the population level, despite being non-identifiable at the individual level.  We illustrate this point numerically in the apprendix, where we illustrate a result of \citet{Lavielle2016} on a simple exponential growth model that provides a classic example of non-identifiability on the individual level.  Motivated by these previous results, we then turn to the Friberg and standard viral dynamics model to illustrate the proposed nonparametric approach to practical identifiability. For both models, we fit a synthetic data set using a multi-start optimisation approach as suggested by \citet{Duchesne2021}, and compare the resulting fits via information criteria. Using the Kolmogorov-Smirnov two-sample test and the overlapping index, we then determine if the population-level parameter distributions are significantly different. We show that circulating neutrophil concentrations from a population of 15 virtual individuals are sufficient to identify the Friberg model at the population level. Conversely, we demonstrate that distinct population-level parameter distributions can result in model fits that equally well describe circulating viral load data. Consequently, we conclude that the standard viral dynamics model is practically unidentifiable, even when considering viral load data from the population level. 

The remainder of the article is structured as follows. 
We next introduce the Friberg and standard viral dynamics models, including the generation of synthetic data from each of these models. We then detail our nonparametric approach to practical identifiability of these models within a NLME framework. We illustrate the approach through two examples before concluding with a brief discussion. 

\section{Methods}

\subsection{Friberg model of chemotherapy-induced neutropenia}

\begin{figure}[!h]
\centering
\includegraphics[trim= 4 5 5 10,clip,width=1\textwidth]{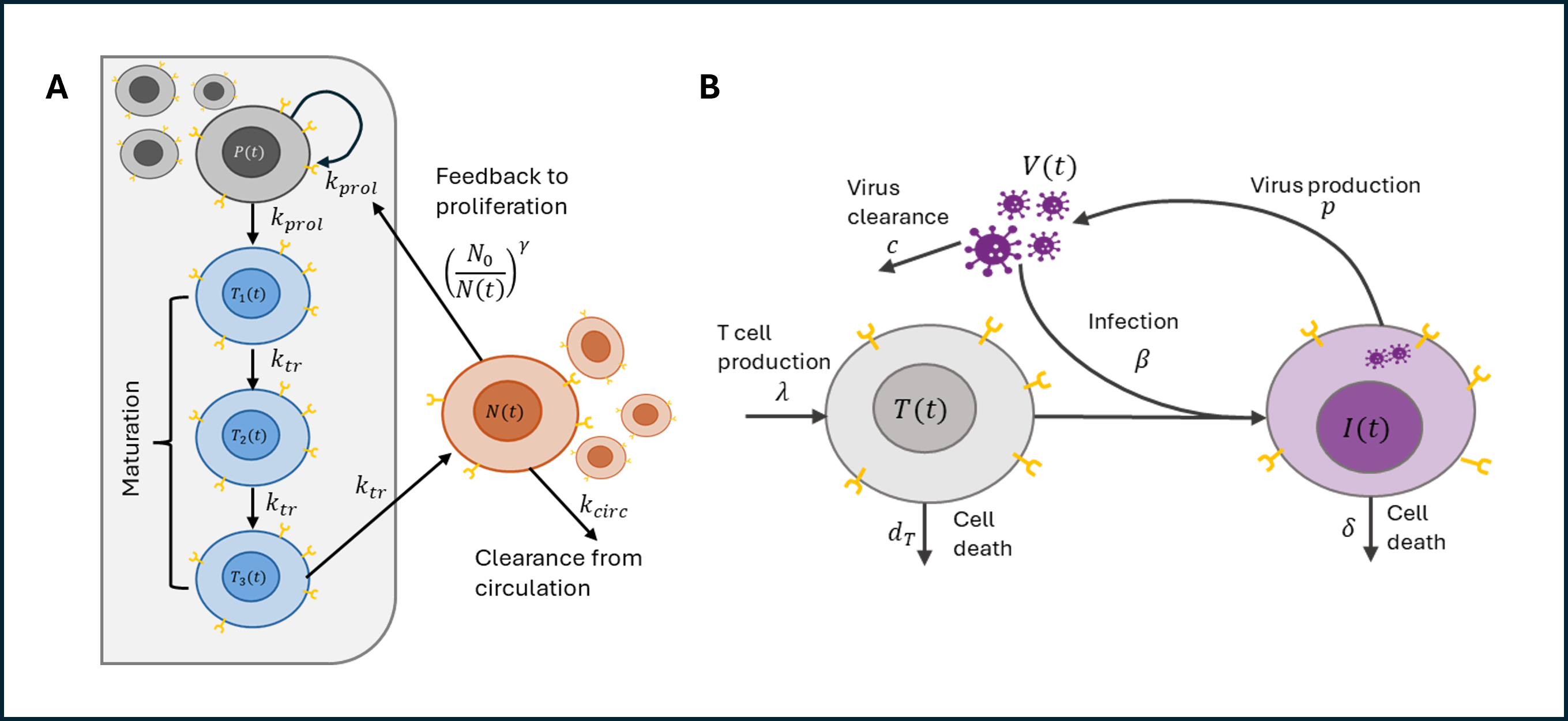}
 \caption{\textbf{Model schematics for the Friberg model of granulopoiesis, and the standard viral dynamics model.} \textbf{A)} Precursor cells, $P$, proliferate within the bone marrow at rate $k_{prol}$ and transit to the first maturation compartment $T_1$ at the same rate. Cells within the maturation compartments $T_{1,2,3}$ progress through the compartments with rate $k_{tr}$ before entering the circulation. Cells within the circulation are cleared at a per capita rate $k_{circ}$ and control the proliferation of precursors through the nonlinear feedback term. 
 \textbf{B)} Uninfected target cells, $T(t)$, are produced at a constant rate $\lambda$ and cleared at a per capita rate $d_T$. Target cells are infected by virus, $V$, with rate constant $\beta $. The resulting infected cells, $I(t)$, produce virus at a per capita rate $p$ and are cleared with rate $\delta.$ Free virus is cleared at per capita rate $c$. }
\label{Fig:ModelDiagram}  
\end{figure}

Neutrophils are the most common white blood cell in humans and an important part of the immune system. Neutropenia, defined as circulating neutrophil cells below a critical level, is a common dose-limiting toxicity of anti-cancer treatments \citep{Cassidy2020b,Craig2017,Friberg2002}. \citet{Friberg2002} developed a semi-mechanistic transit compartmental model of neutrophil production that has become the gold-standard framework for modelling hematopoietic toxicity \citep{Craig2017}. This modelling framework has been extended beyond neutrophils to include other hematopoietic lineages and has been used throughout the pharmaceutical sciences \citep{Craig2017,Soto2020,Quartino2014,deSouza2017}. The \citet{Friberg2002} model includes the dynamics of proliferating neutrophil precursors ($P(t)$), three compartments representing maturation stages in the bone marrow ($T_{1,2,3}(t)$), and a circulating neutrophil concentration that is typically fit to clinical neutrophil measurements ($N(t)$). \citet{deSouza2017} showed that the \citet{Friberg2002} formulation of this model imposes an implicit relationship between the mean maturation time and the proliferation rate ($k_{prol}$) of neutrophil precursors. However, these two processes are mechanistically distinct. Consequently, \citet{deSouza2017} proposed the following generalization of the Friberg model that does not impose this relationship and is given by
 \begin{equation} \label{Eq:FribergODE}
 \left.
 \begin{aligned} 
     \TimeDeriv P(t) & =  \left[ (1-E_{drug}(t))\left( \frac{N_0}{N(t)} \right)^{\gamma} - 1 \right] k_{prol}P(t) \\
     \TimeDeriv T_1(t) & = k_{prol} P(t) - k_{tr}T_1(t) \\
     \TimeDeriv T_{i}(t) & = k_{tr} T_{i-1}(t) - k_{tr}T_i(t) \quad \textrm{for} \quad i = 2,3 \\
     \TimeDeriv N(t) & =  k_{tr} T_3(t) - k_{circ} N(t).
 \end{aligned} 
 \right \}
 \end{equation}
 Eq.~\eqref{Eq:FribergODE} decouples the proliferation rate $k_{prol}$ from the maturation rate $k_{tr}$, while \citet{Friberg2002} makes the modelling assumption that these rates are equal.
 The proliferation of these cells is modulated by the ratio $N_0/N(t)$. This feedback from circulating neutrophils to proliferating neutrophil precursors acts as a cipher for the cytokine effects, and $\gamma$ controls the strength of this feedback.  Neutrophil precursors in the bone marrow progress through the maturation compartments with rate $k_{tr}$. Finally, circulating neutrophils are cleared at a per capita rate $k_{circ}.$ As is commonly done \citep{Friberg2002,deSouza2017}, we take the initial conditions of Eq.~\eqref{Eq:FribergODE} as
\begin{align*}
    N(0) = N_0, \quad  T_{1,2,3}(0) = \frac{k_{circ}N(0)}{k_{tr}},  \quad \textrm{and} \quad P(0) = \frac{k_{circ}N(0)}{k_{prol}}.
\end{align*}
We give a schematic diagram of the model in Eq.~\eqref{Eq:FribergODE} in panel A of Fig~\ref{Fig:ModelDiagram}.

We couple the semi-mechanistic model Eq.~\eqref{Eq:FribergODE} with the four compartment population pharmacokinetic (PK) model for the chemotherapeutic Zalypsis to model chemotherapy-induced neutropenia \citep{Perez-Ruixo2012,Craig2016,Cassidy2020b}, given by 
 \begin{equation}\label{Eq:ZalypsisPK}
 \left.
 \begin{aligned}
     \TimeDeriv C_p(t) & = I_c(t) + k_{fp} C_f(t) + k_{sl1p} C_{sl1}(t) - (k_{pf} + k_{psl1} + k_{cl} ) C_p(t) \\
     \TimeDeriv C_f(t) & = k_{pf}C_p(t) + k_{sl2f} C_{sl2}(t) - (k_{fp} + k_{fsl2})C_f(t)    \\
     \TimeDeriv C_{sl1}(t) & = k_{psl1}C_p(t) - k_{sl1p} C_{sl1}(t) \\
     \TimeDeriv C_{sl2}(t) & = k_{fsl2} C_f(t) - k_{sl2f} C_{sl2}(t).
 \end{aligned}
 \right \}
 \end{equation}
In the PK model in Eq.~\eqref{Eq:ZalypsisPK}, $C_p(t)$ represents drug concentration in the central compartment, $C_f(t)$ is the concentration in the fast-exchanging tissues, and $C_{sl1}(t)$ and $C_{sl2}(t)$ are the concentrations in the slow-exchange tissues. The rates $k_{ij}$ give the transit rate from compartment $i$ to compartment $j$ while $I_c(t)$ models the subcutaneous administration of Zalypsis. We model the effect of chemotherapy on blocking the effective proliferation of neutrophil precursors, $P(t)$, using the standard Emax model
\begin{align*}
    1-E_{drug}(t) = 1- \frac{E_{max}C_p(t)}{EC_{50} + C_p(t)},
\end{align*}
where $E_{max}$ is the maximal inhibition of proliferation of neutrophil precursors and $EC_{50}$ is the concentration that results in half-maximal inhibition of neutrophil proliferation. 

\subsubsection{Prior identifiability results on Friberg model}

\citet{Evans2018} studied the structural identifiability of a class of semi-mechanistic models related to the Friberg model. There, they found that, if the initial conditions of the Friberg model correspond to equilibrium, as is commonly done, and $k_{tr} = k_{prol}$, then the model is structurally identifiable. However, if both assumptions are not met, the model is structurally unidentifiable. 

As discussed by \citet{deSouza2017}, the assumption $k_{tr} = k_{prol}$ imposes a relationship between two physiologically distinct processes, namely the maturation rate of neutrophil precursors, and the proliferation rate of proliferating cells. As there is no \textit{a priori} biological reason to assume that these rates are identical, we do not impose $k_{tr} = k_{prol}$. Consequently, the results of \citet{Evans2018} indicate that the model parameters are structurally unidentifiable at the individual level. 

\subsubsection{Generating a virtual population and synthetic data}

We generated a virtual population using the population parameter distributions for docetaxel treatment that are reported in Table 4 of \citet{Friberg2002}. The population parameter estimates defining the neutrophil dynamics and antiproliferative effects of the anticancer treatment were taken from \citet{Friberg2002} and are therefore self-consistent. We included inter-individual variability on model parameters corresponding to the baseline neutrophil concentration, $N_0$, the $EC_{50}$, the maturation rate $k_{tr}$, and, as we are not enforcing $k_{tr} = k_{prol},$ we include inter-individual variability on $k_{prol}$. We assumed that all parameters with inter-individual variability followed a log-normal distribution at the population level. Here, we follow \citet{Friberg2002} and log-transform the inter-individual variability sampled from Normal distributions. As our model formulation slightly differs from \citet{Friberg2002}, we chose the standard deviations of the inter-individual variability to ensure that the simulated neutrophil dynamics reflect the heterogeneity in Figure~1 of \citet{Friberg2002}. The mean and variance of these population distributions are given in Table~\ref{Table:NeutropeniaKnownParameter}.
\begin{table}[!ht]
\begin{tabular}{llll} 
Parameter (units)   & Population estimate & Standard deviation  & Distribution  \\  \hline
$\log(N_0)$ ($10^9$ cells/mL)  & $\log(5.03)$  & 0.22 & Normal \\ \hline
$\log(EC_{50})$ (ng/mL)  &  $\log(0.14)$  & 0.33  & Normal \\ \hline
$\log(k_{tr})$  (1/day)  & $\log(1.08)$ & 0.41  & Normal \\ \hline
$\log(k_{prol})$ (1/day)  & $\log(0.87)$ & 0.42  & Normal \\ \hline
$\gamma$ (unitless)  & 0.16 & N/A & N/A \\ \hline
$k_{Circ}$ (1/day)  & 1.15 & N/A & N/A \\ \hline
$E_{Max}$ (unitless)  & 1 & N/A & N/A \\ \hline 
  \\  
\end{tabular}
\caption{ \textbf{Population parameters used to generate simulated data using the Friberg neutropenia model.} A standard deviation of N/A indicates that the given parameter was fixed across the entire virtual population. The population parameter estimates are taken from Table 4 of \citet{Friberg2002}. Consistent with the parameter estimation in \citet{Friberg2002}, we sample from normal distributions before transforming the parameter value to the linear scale. }
\label{Table:NeutropeniaKnownParameter}
\end{table}

We utilized the population PK parameter estimates from \citet{Perez-Ruixo2012} to parameterize the PK model in Eq.~\eqref{Eq:ZalypsisPK}. As we are primarily interested in the identifiability of the Friberg model of chemotherapy-induced neutropenia, we do not include any inter-individual variability on the PK parameters. We generated synthetic circulating neutrophil concentrations for $N= 15$ virtual participants, by sampling from the known population parameter distributions in Table~\ref{Table:NeutropeniaKnownParameter}, reflecting the mean size of phase I trials of docetaxel \citep{Clarke1999}.
For each virtual participant, we simulate the Friberg model in Eq.~\eqref{Eq:FribergODE} for 65 days with Zalypsis administrations on days 0, 21, 42, and 63. We measure the corresponding neutrophil concentration on days $t_j = 0,3,6,9,...,63$. 

As mentioned above, the Friberg model is structurally unidentifiable at the individual level. We wish to test whether simultaneously fitting multiple individuals receiving the same treatment will allow for accurate estimation of the population parameter distributions, even if the underlying individual model is unidentifiable. In what follows, we consider the idealized situation, where the neutrophil measurements are noise-free. This implies that the model is perfectly specified to the experimental data. We can therefore study the identifiability of the model parameters without considering any model misspecification or measurement error \citep{Browning2024}.

\subsubsection{Parameterization of the Friberg model}

To fit the synthetic neutrophil data, we first fixed the parameters defining the PK model in Eq.~\eqref{Eq:ZalypsisPK} to existing population parameter estimates \citep{Craig2016,Cassidy2020b}. Moreover, we fixed the maximal anti-proliferative effect of chemotherapy as $E_{max} = 1$. We further assumed that the clearance rate of circulating neutrophils, $k_{circ},$ and the strength of feedback from circulation to proliferation, $\gamma$, are approximately equal between participants in this simulated clinical trial. Thus, we estimated $k_{circ}$ and $\gamma$ at the population level without inter-individual variability. We allowed the remaining parameters, $k_{prol}, N_0, EC_{50},$ and $k_{tr}$ to have inter-individual variability. 

We fit the Friberg model~\eqref{Eq:FribergODE} to the synthetic data using the nonlinear mixed-effects framework implemented in Monolix using the stochastic approximation expectation maximization algorithm (SAEM) \citep{Lavielle2014,Monolix2020,Kuhn2005,Delyon1999}. We assumed that the model parameters with inter-individual variability, $\phi = [k_{prol}, N_0, EC_{50},k_{tr}]$, follow a log-normal distribution at the population level. The population parameter distribution is defined by 
\begin{equation*} 
    \phi_i = \varphi_i  e^{\psi_i} ,	
\end{equation*}
where $\varphi_i$ is the population estimate, $\omega$ is the standard deviation of the random effects, and $\psi_i \sim \mathcal{N}(0, \omega_i^2)$ captures the inter-individual variability for each parameter \citep{Monolix2020,Lavielle2014}.

We fit the Friberg model in Eq.~\eqref{Eq:FribergODE} to the synthetic neutrophil data sampled at $ 22$ distinct time points $t_{j} = 0,3,6,9,...,63$ for the $N = 15$ virtual participants. We denote the synthetic data from the $i$th virtual participant at time $t_j$ by $y_{i,j}$. Now, let $\theta_i$ be the model parameters for the $i$th virtual participant and $f(t_{j} ,\theta_{i})$ be the simulated neutrophil concentration, $N(t_j),$ for model parameters $\theta_i$. For each participant, the statistical model is therefore given by
\begin{equation*} 
    y_{i,j} = f(t_{j} ,\theta_{i}) + g(f(t_{j} ,\theta_{i}) , \omega
    )  e_{i,j} .	
\end{equation*} 
In this framework, the error model is given by $g(f(t_{j} ,\theta_{i}) , \omega ) e_{i,j} $, where $e_{i,j}$ is the residual error of participant $i$ at measurement $j$, and $\omega$ is a vector of error parameters. The residual error, $e_{i,j}$, is assumed to follow a normal distribution with mean 0 and variance $1$, so the error function $g(f(t_{j} ,\theta_{i}) , \omega )$ determines the standard deviation of the residual error model \citep{Lavielle2014}. For the synthetic neutrophil concentrations, we consider a proportional error model, so 
\begin{align*}
    g (f(t_{j} ,\theta_{i}) , \omega ) = b[f(t_{j},\theta_i)] ,
\end{align*}
where $b$ is an error parameter that is estimated during parameter fitting. 

We estimated the model parameters using the SAEM algorithm implemented in Monolix \citep{Kuhn2005,Monolix2020}. Even in the case of identifiable parameters, the SAEM algorithm may converge to a local, rather than global, optimum of the likelihood. Therefore, we followed the multi-start procedure suggested by \citet{Duchesne2021} and generated 100 different initial estimates for the model parameters by randomly sampling in biologically plausible ranges. We then estimated the model parameters from each of these 100 initial estimates, which corresponds to the multi-start approach suggested by \citet{Duchesne2021}. We compared the goodness-of-fit for each of the 100 posterior parameter estimates using the Akaike Information Criteria (AIC). As the model structure and number of parameters were fixed, this is equivalent to comparing parameter estimates by the corresponding loglikelihood.  

\subsection{Standard viral dynamics model}

The standard viral dynamics model has been extensively used to understand the dynamics of viral infection in HIV-1 and other viral infections \citep{Perelson2002,Perelson1996,Hill2018,Smith2011b,Mainou2024}. In the most common formulation, the model tracks the concentration of uninfected target cells, $T(t)$, infected cells $I(t)$, and free infectious virus $V(t)$. Here, we consider HIV-1 dynamics, where the target cells represent CD$4^+$ T-cells. We assume that these cells are produced at a constant rate $\lambda$ and cleared at a per capita rate $d_T$. Infection occurs at a rate $\beta$ following contact between a target cell and infectious virion, $V(t)$, and these infected cells are cleared at per capita rate $\delta$. Infected cells produce virus at a constant rate $p$ and this free virus is cleared from the circulation at a per capita rate $c$. All told, the model is given by 
\begin{equation}\label{Eq:TIVDynamicsModel}
\left. 
\begin{aligned}
    \TimeDeriv T(t) &= \lambda - \beta V(t)T(t) - d_T T(t), \\
    \TimeDeriv I(t) &= \beta V(t)T(t) - \delta I(t), \\
    \TimeDeriv V(t) &= p I(t) - cV(t). 
\end{aligned}
\right \}
\end{equation}
The system in Eq.~\eqref{Eq:TIVDynamicsModel} is equipped with initial conditions $T(0) = T_0, I(0) =I_0, $ and $V(0) = V_0$, and illustrated in Panel B of Figure~\ref{Fig:ModelDiagram}.

\subsubsection{Prior structural identifiability results on the standard viral dynamics model}
 The standard viral dynamics model has been extensively studied. \citet{Miao2011} and \citet{Wu2008} show that the viral production rate $p$ cannot be typically identified from viral load data alone. 

\subsubsection{Generating a virtual population and synthetic data}

Here, we consider the dynamics of the concentrations of circulating viral loads after acute HIV-1 infection. To generate synthetic viral load data, we sampled model parameters from population distributions to generate virtual patients. We did not obtain these population-level parameters by fitting the model to clinical data. However, these population values are broadly consistent with estimates in the literature \citep{Conway2015,Phan2024a,Hill2018,Luo2012,Catalfamo2008} and are given in Table~\ref{Table:ViralDynamicsKnownParameter}. 
\begin{table}[!ht]
\begin{tabular}{llll} 
Parameter (units)   & Population estimate & Standard deviation & Distribution \\  \hline
$\log (\beta)$ (mL/copies/day )  & $\log(8 \times 10^{-7} )$ & 0.35 & Normal \\ \hline
$\log (p) $  (copies/cell/day)  & $\log(3500) $ & 0.4 & Normal \\ \hline
$\log( \delta ) $ (1/day)  & $\log( 0.25)$  & 0.35 & Normal \\ \hline
$\log (T_0) $ (cells/mL)  &  $\log(1.5 \times 10^6 ) $ & 0.45 & Normal \\ \hline
$\log_{10} (V_0)$ (copies/mL)  & $1 \log_{10} $ & 0.25 & Normal \\ \hline 
\end{tabular}
\caption{ \textbf{Population parameters used to generate simulated data using the standard viral dynamics model.} Consistent with the parameter estimation in Monolix, we sample from normal distributions before transforming the parameter value to the linear scale.  }
\label{Table:ViralDynamicsKnownParameter}
\end{table}

To determine the initial conditions of the model, we recall that we are considering acute infection. We consider viral load data beginning with the first quantifiable circulating viral load measurement as $V(0).$ As is typical when modelling viral loads measured in clinical studies, we consider the $\log_{10}$ transformed circulating viral load \citep{Cassidy2023,Ismail2021,Reeves2020,Phan2024a,Stephenson2021,Reeves2023}. Consequently, we assume that the first quantifiable viral load measurement is normally distributed with mean $ 1\log_{10}$ copies/mL. This baseline concentration is consistent with the lower limit of quantification of highly sensitive assays \citep{Palmer2003}. 

We fix the remaining model parameters at commonly accepted values \citep{Cassidy2023,Reeves2020,Lim2024,Cassidy2023a}. We set the infection-independent death rate of target cells to $d_T  = 0.01$ day$^{-1}$ \cite{Mohri2001}. Now, we assume that the concentration of CD4 T-cells is approximately at the uninfected steady state, so we calculate the production rate of target cells via
\begin{align*}
    \lambda = T(0) d_T.
\end{align*}
We fix the clearance rate of viral particles at $c = 23$ day$^{-1}$ \cite{Ramratnam1999}, which implies that the concentration of infected cells at $t = 0$ is given by 
\begin{align*}
    I(0) = cV(0)/p.
\end{align*}
Once again, we considered a virtual population reflecting the typical size of phase I trials of anti-HIV antivirals \citep{Stephenson2021,Mendoza2018,Sneller2022}. 
We generate synthetic viral load data by sampling $N = 15$ virtual participants from the known population parameter distributions. For each virtual participant, we simulate the viral dynamics model in Eq.~\eqref{Eq:TIVDynamicsModel} for 65 days. We measure the $\log_{10}$ viral load concentrations on days $t_j = 0,8,12,16,...,64$ and add IID Gaussian measurement error with mean $0$ and standard deviation $\sigma = 0.1 \log_{10} $ copies/mL. Here, unlike the previous example, we consider the more realistic scenario, where the viral load measurements are not noise-free. Consequently, we study the identifiability of the standard viral dynamics model in the case where there is potential model misspecification or measurement error.
 
\subsubsection{Parameterization of the standard viral dynamics model}

As mentioned, the death rate of uninfected CD4 T-cells is typically set to $d = 0.01$/day and the per capita clearance rate of free virus is taken as $c = 23$/day. Consequently, we fix these parameters when fitting the synthetic viral load data. We assumed that $V(0)$ represents the first quantifiable viral load measurement. However, this viral load measurement results from an infection event at time $t_0 < 0$. We therefore estimated  the concentration of CD4 T-cells, $T_0 = T(0)$, and viral load, $V_0 = V(0)$, at time $t = 0$. 

As before, we estimated the model parameters using the nonlinear mixed-effects framework implemented in Monolix \citep{Lavielle2014,Monolix2020}. Here, we assume that the model parameters $\phi = [\beta, p, \delta, T_0, V_0]$ follow a log-normal distribution at the population level defined by 
\begin{equation*} 
    \phi_i= \varphi_i  e^{\psi_i} ,	
\end{equation*}
where $\varphi_i$ is the population estimate and $\psi_i \sim \mathcal{N}(0, \omega^2)$ captures the inter-individual variability \citep{Monolix2020,Lavielle2014}. We included inter-individual variability for all estimated parameters.

We fit the standard viral dynamics model Eq.~\eqref{Eq:TIVDynamicsModel} to synthetic HIV viral load that was sampled at $16$ distinct time points, $t_{j} = 0,8,12,16,...,64$. For the $i$th virtual participant, the $\log_{10}$ synthetic viral load concentration at time $t_j$ is denoted by $y_{i,j}$, while the vector of model parameters for the $i$th individual is given by $\theta_{i}$. As in the case of the Friberg model, the statistical model is given by
\begin{equation*} 
    y_{i,j} = \log_{10}( f(t_{j} ,\theta_{i}) ) + g(f(t_{j} ,\theta_{i}) , \omega )  e_{i,j} ,	
\end{equation*}     
where $f(t_{j} ,\theta{i})$ is the solution $V(t_j)$ of Eq.~\eqref{Eq:TIVDynamicsModel}. There are many common choices for the error function $g(f(t_{j} ,\theta) , \omega )$ in viral dynamics modelling \citep{Goncalves2021,Zitzmann2023,Cao2021,Clairon2023,Marc2023,Iyaniwura2024}. We are estimating the $\log_{10}$ viral load concentration, so we set
\begin{equation*} 
    y_{i,j} =  \log_{10}( f(t_{j} ,\phi_{i}) ) + a e_{i,j}.  
\end{equation*}  
Therefore, the error model scales the residual error $e_{i,j}$ by the error parameter $a$, which we estimate during parameter fitting. 

As before, we estimated the model parameters using the SAEM algorithm implemented in Monolix \citep{Kuhn2005,Monolix2020}. We once again considered 100 different initial estimates for the model parameters by randomly sampling in biologically plausible ranges, which corresponds to the multi-start approach suggested by \citep{Duchesne2021}. We compared the goodness-of-fit for each of the 100 posterior parameter estimates using the AIC. 

\subsection{Analysis of population parameter distributions}
We analyse model fits that capture the underlying synthetic data. For both the Friberg and viral dynamic models, we consider the ten best fits to the synthetic data as measured by AIC. In both these examples, all model fits represent very good fits to the synthetic data, so our choice to compare the ten best is somewhat arbitrary. Now, as we are only considering fits that capture the synthetic data, any significant differences between the estimated parameter distributions would, therefore, indicate nonidentifiability. However, variation between the resulting estimates for the parameter populations is to be expected, as we estimate the population-level distribution by fitting data from a finite sample of individuals from the population. Consequently, rather than directly comparing the parametric representations of the population distributions, as proposed by \citep{Duchesne2021}, we will utilize statistical techniques to determine if these differences are significant. 
 
We consider $j = 10$ population distributions for each parameter and $N=15$ individual parameter estimates corresponding to the $N$ virtual participants. Let $\{\phi_{i,j}\}_{i=1}^N$ be the $N$ individual parameter estimates for the $i$th parameter of the $j$th fit, and denote the population distribution of the $i$th parameter from the $j$th fit by $\Phi_{i,j}$. The individual estimates $\{\phi_{i,j}\}_{i=1}^N$ correspond to $N$ samples from the underlying population distribution, $\Phi_{i,j}$.  As we consider nonparametric approaches, we do not use the parametric representation of $\Phi_{i,j}.$ 
 
 We begin by evaluating if the $N$ individual estimates, $\{\phi_{i,j}\}_{i=1}^N$, are sufficient to distinguish between the underlying population distributions from two distinct fits. The Kolmogorov-Smirnov two-sample test is a nonparametric test to assess if these $N$ samples were taken from different underlying population distributions. Thus, we consider the individual estimates from each possible pair of distinct fits, $\{\phi_{i,j}\}_{i=1}^N$ and $\{\phi_{i,k}\}_{i=1}^N$, and use the Kolmogorov-Smirnov test to identify significant differences between $\Phi_{i,j}$ and $\Phi_{i,k}$.  We use the significance threshold $\alpha = 0.05$ to reject the null-hypothesis that the samples are drawn from the same underlying distribution. However, the Kolomogorov-Smirnov test relies on the ability to well-order the individual estimates $\Phi_{i,j}$. Consequently, when considering correlated parameters, we consider the nonparametric maximum mean discrepancy test \citep{Gretton2012,Gao2023} to test if two multivariate samples are drawn from different underlying distributions.

We also seek to compare the parameter estimates at the population level. To do so, we use the overlapping index \citep{Pastore2019,Schmid2006}. We consider all pairs of distinct population distributions of the $i$th parameter, $\Phi_{i,j}$ and $\Phi_{i,k}$, with corresponding probability density functions $f_{i,j}$ and $f_{i,k}$. The overlapping index between $\Phi_{i,j}$ and $\Phi_{i,k}$ is given by  
 \begin{align*}
     o_{i,j,k} = \int_{-\infty}^{\infty} \min\left( f_{i,j}(\theta),f_{i,k}(\theta) \right) \d \theta. 
 \end{align*}
An overlapping index of $o_{i,j,k} = 1$ indicates that the two probability measures are equal almost-surely. Conversely, $o_{i,j,k} = 0 $ implies that the two distributions are entirely disjoint. The overlapping index has recently been used to distinguish between responders and non-responders in a virtual clinical trial \citep{Mongeon2025}. Essentially, this nonparametric approach measures the agreement between the two probability distributions and is directly related to the total variation distance between the corresponding measures. The overlapping index immediately extends to the multivariate case by calculating a higher-dimensional integral and facilitates the comparison of multivariate population parameter distributions.

Finally, model parameters that were estimated at the population-level, i.e. without inter-individual variability, correspond to point-estimates. Consequently, studying the identifiability of these parameters without inter-individual variability does not require understanding if two distributions are significantly different, but rather studying if the data is sufficient to constrain the point-estimate. Therefore, the identifiability of these point-estimates can be directly studied using existing techniques focused on individual-level dynamics such as profile likelihood analysis \citep{Raue2009}. 
\section{Results}

\subsection{The Friberg model of chemotherapy-induced neutropenia is identifiable at the population level}
We now consider the model fits to the chemotherapy-induced neutropenia synthetic data using the Friberg model in Eq.~\eqref{Eq:FribergODE}. Here, we consider the 10 best fits to the data as measured by the AIC. The corresponding log-likelihood values for these fits are shown in Table~\ref{Table:NeutropeniaGoodnessOfFitResults}.   
\begin{table}[!ht]
\begin{tabular}{ll|ll|ll} 
Fit Number  & $-2\mathcal{LL}$ & Fit Number  & $-2\mathcal{LL}$ & Fit Number  & $-2\mathcal{LL}$ \\  \hline  
Fit 1 & -1392.4 & Fit 4 & -1029.6 & Fit 8 & -797.4  \\ \hline
Fit 2 & -1264.1 & Fit 5 & -972.4 & Fit 9 & -727.7 \\ \hline
Fit 3 & -1251.1 & Fit 6 & -933.3 & Fit 10 & -725.6 \\ \hline 
& & Fit 7 & -916.4 & &  \\ \hline
\end{tabular}
\caption{\textbf{Summary of the 10 best fits to the synthetic neutrophil concentration data}. The 10 best fits to the synthetic neutrophil concentration data using Eq.~\eqref{Eq:FribergODE} as measured by the log-likelihood. Each column, corresponding to Fits (1-3), (4-7), and (8-10), represent distinct local maxima of the log-likelihood. }
\label{Table:NeutropeniaGoodnessOfFitResults}
\end{table}

\begin{figure}[!h]
\centering
\includegraphics[trim= 4 5 5 10,clip,width=1\textwidth]{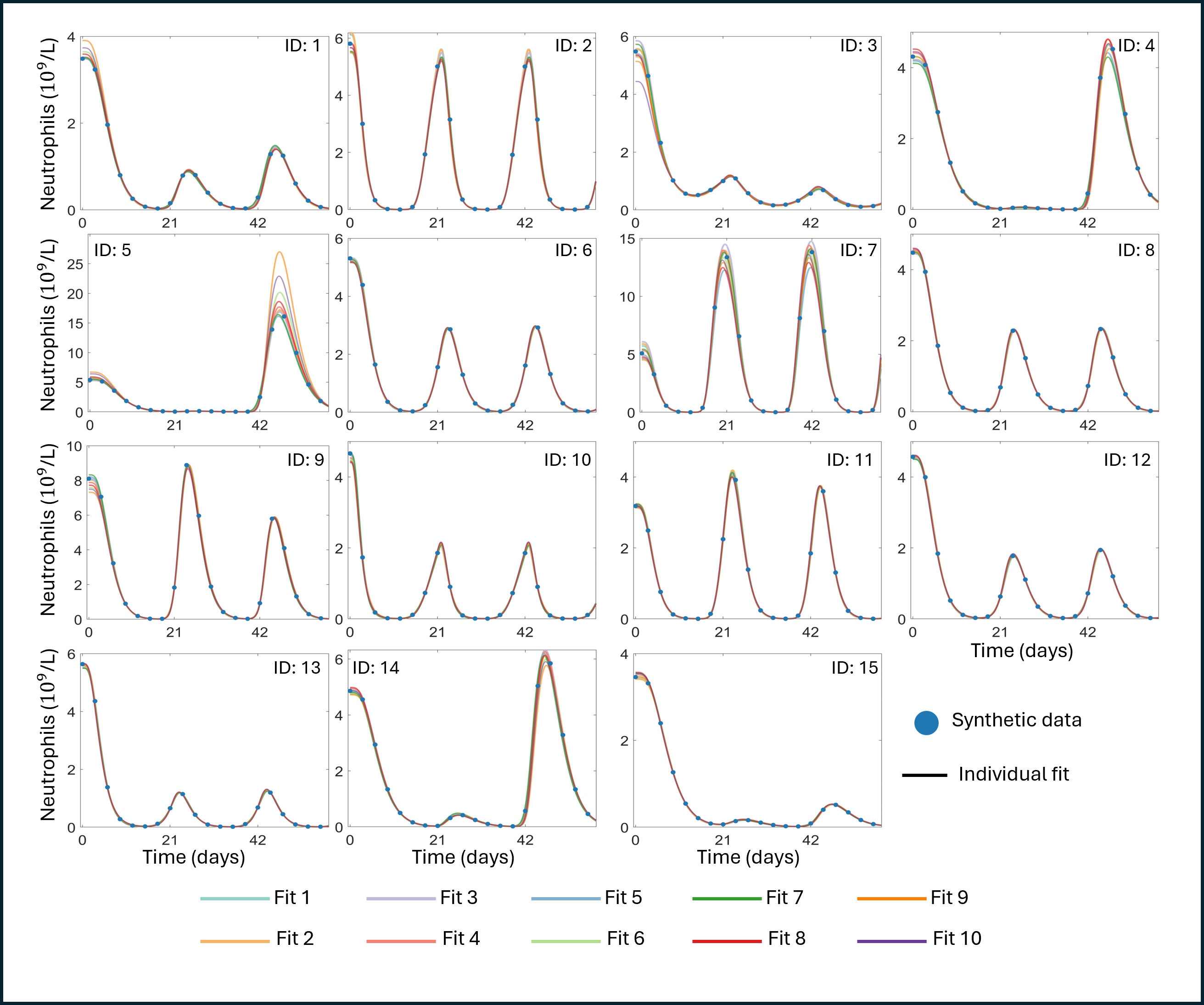}
 \caption{ \textbf{A comparison of the 10 best fits to the individual synthetic neutrophil concentration data}. The individual fits to the viral load data from 15 virtual patients using Eq.~\eqref{Eq:FribergODE}. In all cases, the synthetic neutrophil concentrations are shown as solid blue circles. The fit circulating neutrophil concentrations for each of the virtual participant in each of the 10 best fits are shown as solid lines, distinguished by color. }
\label{Fig:FribergIndividualFits}  
\end{figure}

The goodness-of-fits for these 10 best fits identify three distinct local maxima of the log-likelihood, at $-2\mathcal{LL} \approx -1250, -950$ and $-790.$ Despite these differences, the model fits the synthetic data quite well for each of the 10 best fits. Now, we considered a large number of fits from throughout the plausible parameter range. Therefore, we conclude that fits 1-3 from Table~\ref{Table:NeutropeniaGoodnessOfFitResults}, with the highest log-likelihood values, represent the global maximum of the log-likelihood, while the other two clusters represent local maxima. For each of the 10 best fits, we show the individual fits to the synthetic circulating neutrophil concentrations in Figure~\ref{Fig:FribergIndividualFits}.  For some individuals, the predicted dynamics are identical across all 10 best fits. However, some individual fits illustrate subtle differences in the predicted neutrophil dynamics during chemotherapy. Specifically, for individuals with oscillatory neutrophil dynamics with a relatively large amplitude (ID: 4, 5, 7, 9, 14), there is a discernible difference between the model predictions across the 10 best fits. These differences are reflected in the corresponding log-likelihood value for each of these fits as shown in Table~\ref{Table:NeutropeniaGoodnessOfFitResults}.

\begin{figure}[!h]
\centering
\includegraphics[trim= 4 5 5 10,clip,width=1\textwidth]{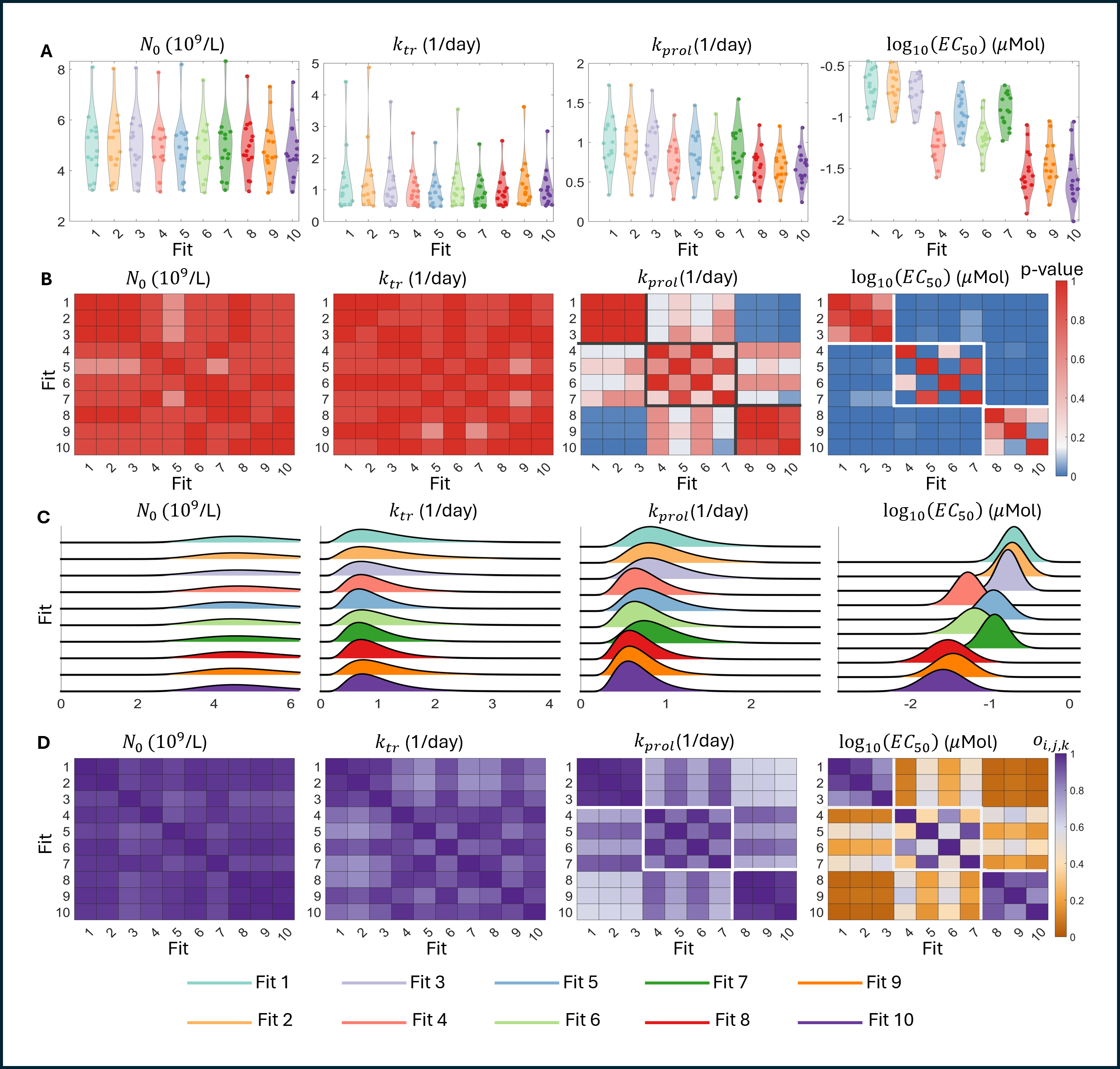}
 \caption{\textbf{Assessing practical identifiability of the Friberg model}.  Panel A shows the distribution of the 15 individual estimates for the model parameters $N_0, k_{tr}, k_{prol}$ and $\log (EC_{50})$ for each of the 10 best population fits as dots within each violin plot. Panel B shows the $p$-value corresponding to the pairwise Kolmogorov-Smirnov test between all 10 best fits for each of the four model parameters, with the grey and white borders identifying the three distinct clusters. Here, $p$-values below the significance threshold of $0.05$ are plotted as blue squares. Panel C shows the 10 estimated population distributions for each of the four model parameters. Panel D shows the overlapping index between each of the pairwise comparisons of the 10 best fits. Here, for each of the $k = 1,2,3,4$ model parameters, pairs of fits $(i,j)$  with an overlapping index $ o_{i,j,k} > 0.5$ are represented by the shades of violet, while pairs with $o_{i,j,k} < 0.5$ are represented by shades of orange and the three clusters are again highlighted by white boxes. }
\label{Fig:FribergIdentifiability}  
\end{figure}

Having established that the Friberg model is able to recapture the synthetic data, we next turn to the estimated model parameters. In Figure~\ref{Fig:FribergIdentifiability}A, we show the distribution of the 15 individual estimates for each of the 10 best fits. For the model parameters $N_0$ and $k_{tr}$, the distributions of the individual estimates are similar between all 10 best fits. In contrast, for $k_{prol}$ and $\log(EC_{50}),$ there is a clear clustering of the individual distributions of these parameters for fits (1-3), (4-7), and (8-10), which correspond to the three local maxima of the log-likelihood. As we show in Figure~\ref{Fig:FribergIdentifiability}B, the Kolmogorov-Smirnov test did not identify any significant pairwise differences between the individual estimates for the 10 best fits of the model for $N_0$ and $k_{tr}$. However, there are significant differences, measured via the Kolmogorov-Smirnov test, for the parameters $k_{prol}$ and $\log(EC_{50})$ corresponding to the three local maxima. In Figure~\ref{Fig:FribergIdentifiability}B, we highlight these clusters of individual estimates. Importantly, while we do identify significant differences between the individual estimates for these parameters at each local maxima, the estimates within each cluster do not significantly differ from each other, which suggests that these are at least three distinct local optima of the log-likelihood; we can identify either 3 local optima by using the clusters shown in Figure~\ref{Fig:FribergIdentifiability}, or 4 local optima by dividing the second cluster into two subgroups of fits 4 and 6 and fits 5 and 7. Specifically, there are significant pairwise differences between the individual estimates for $\log_{10}(EC_{50})$ these subgroups. However, the population distributions for these parameters display significant overlap, which implies that the 15 individuals considered in this example may not be sufficient to distinguish between the population estimates for this parameter. 

Analysis of the population parameter distributions confirms our observations from the individual parameter estimates. Once again, the population distributions of $N_0$ and $k_{tr}$ are indistinguishable at the population level. Unsurprisingly, these distributions have noticeable overlap shown in Figure~\ref{Fig:FribergIdentifiability}C. Indeed, the mean overlapping indices for the population parameter distributions of $N_0$ and $k_{tr}$ are 0.91 and 0.82, respectively. Conversely, the population distributions for $k_{prol}$ and $\log(EC_{50})$ reflect the previously identified local maxima of the log-likelihood. This clustering is, once again, particularly apparent between fits 1-3 and fits 8-10. There is almost no overlap between these two clusters of population parameter distributions, as shown in Figure~\ref{Fig:FribergIdentifiability}D. However, as before, the population distributions within each of these two clusters do exhibit a high-level of overlap, where the three clusters are highlighted by a white border. Taking the individual and population level results together, we therefore conclude that these three clusters represent distinct maxima of the log-likelihood, but that the fits within each cluster are not significantly different. Thus, our approach for comparing parameter estimates at both the individual and population level is able to distinguish between distinct local maxima of the log-likelihood, but does not identify artificial differences between the parameter estimates within these clusters. We therefore conclude that the Friberg model is practically identifiable at the population level. 

We next compared the estimates for the parameters considered without inter-individual variability, $\gamma$ and $k_{circ}$, against the true values for these parameters for the cluster of best-fits (1-3) . While we did not precisely estimate the known values, this fitting approximately recaptured the underlying truth values with $\gamma= 0.15, 0.15,$ and $\gamma = 0.18$ for fits 1-3, respectively compared against the true value of $\gamma = 0.163$. Similarly, the estimates for $k_{circ}= 0.87$/day, $0.81$/day, and $k_{circ} =1.41$/day compared against $k_{circ} =1.15$/day We also compared the population parameter distributions from fits 1-3 against the known distributions for the parameters with inter-individual variability in Figure~\ref{Fig:ComparisonFribergIdentifiabilitySI}. In general, the population estimates recaptured the underlying known parameter distributions. 
 
\subsection{The standard viral dynamics model is not identifiable at the population level}

We next turn to the model fits to the synthetic HIV viral load using Eq.~\eqref{Eq:TIVDynamicsModel}. We once again select the 10 best fits to this synthetic data, as measured by AIC, with the population log-likelihood given in  Table~\ref{Table:HIVGoodnessOfFitResults}.
\begin{table}[!ht]
\begin{tabular}{ll|ll|ll|ll|ll}  
Fit  & $-2\mathcal{LL}$  & Fit  & $-2\mathcal{LL}$ & Fit  & $-2\mathcal{LL}$  & Fit  & $-2\mathcal{LL}$  & Fit   & $-2\mathcal{LL}$  \\  \hline
Fit 1 & -191.9  & Fit 2 & -191.6 & Fit 3 & -191.6  & Fit 4 & -191.6 & Fit 5 & -191.5 \\ \hline
 Fit 6 & -191.5  & Fit 7 & -191.5 & Fit 8 & -191.5  & Fit 9 & -191.5 & Fit 10 & -191.4  \\ \hline
\end{tabular}
\caption{ \textbf{Summary of the 10 best fits to the synthetic viral load data}. The 10 best fits to the synthetic viral load data using Eq.~\eqref{Eq:TIVDynamicsModel} as measured by the log-likelihood. Here,  lower values of  $-2\mathcal{LL}$  correspond to a better fit.  }
\label{Table:HIVGoodnessOfFitResults}
\end{table}

\begin{figure}[!h]
\centering
\includegraphics[trim= 4 5 5 10,clip,width=1\textwidth]{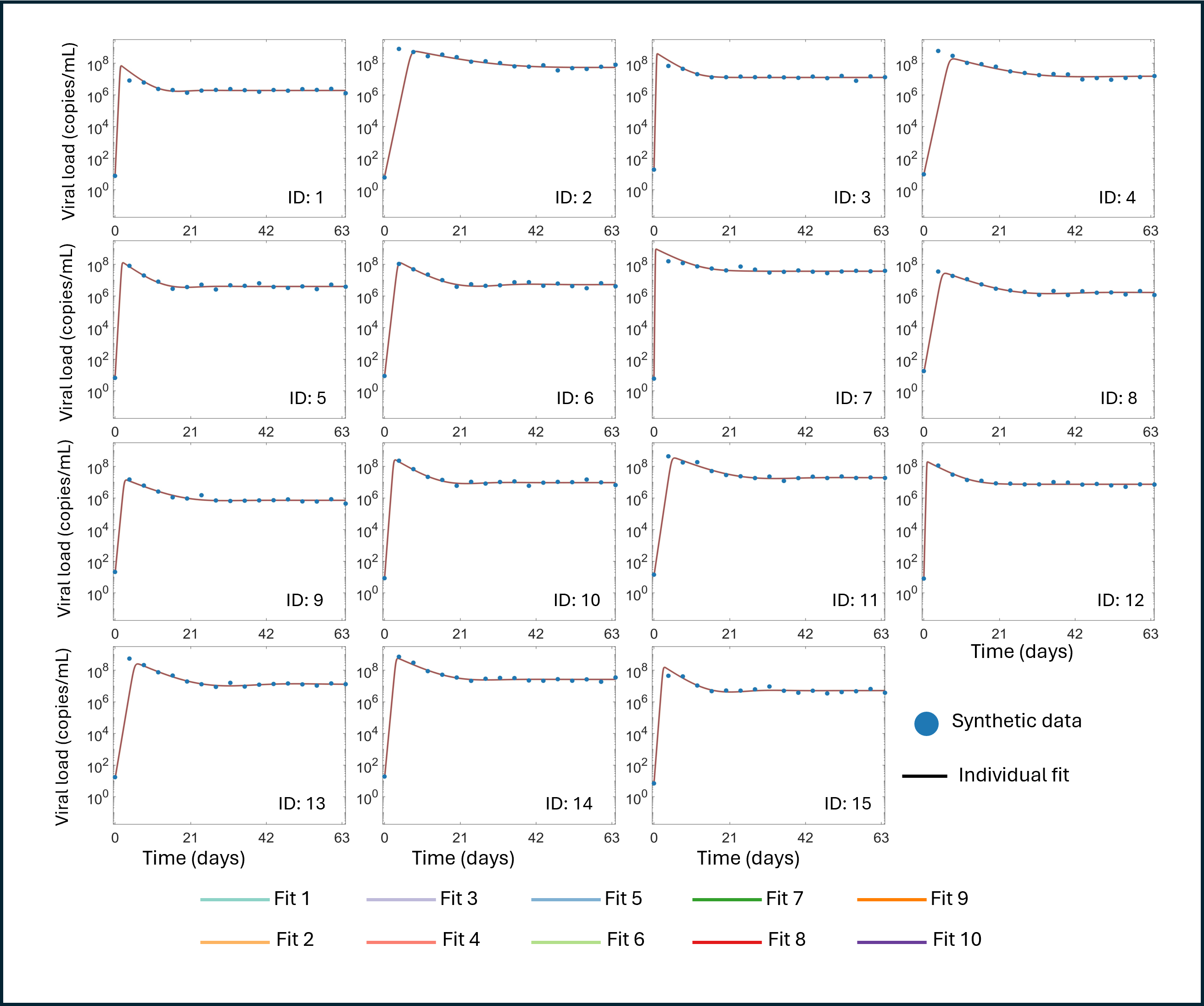}
 \caption{\textbf{A comparison of the 10 best fits to the synthetic viral load data.} The individual fits to the viral load data from 15 virtual patients using Eq.~\eqref{Eq:TIVDynamicsModel}. In all cases, the synthetic viral load data is shown as solid blue circles. The fit circulating viral load for each of the virtual participants resulting from each of the 10 best fits are shown as solid lines, which distinguished by their color, but entirely overlapping.  }
\label{Fig:HIVIndividualFits}  
\end{figure}

We note that all 10 best fits to the synthetic viral load data have extremely similar log-likelihood values, which indicates no appreciable differences between the best-fit and the 10th best fit. Indeed, the worst of the 100 distinct parameterizations of this model has a  $-2\mathcal{LL}$ value of -189.95, which is comparable to the best fit. Consequently, we conclude that these 10 best fits have converged to the maximum of the log-likelihood. In Figure~\ref{Fig:HIVIndividualFits}, we show the individual fits to each of the 15 virtual participants from all 10 best fits. As we would expect from the log-likelihood, the model fits the synthetic data well for each of the 10 best fits. Furthermore, the predicted individual dynamics are indistinguishable for all 15 virtual participants. We therefore conclude that the model can recapture the synthetic data.  

\begin{figure}[!h]
\centering
\includegraphics[trim= 4 5 5 10,clip,width=1\textwidth]{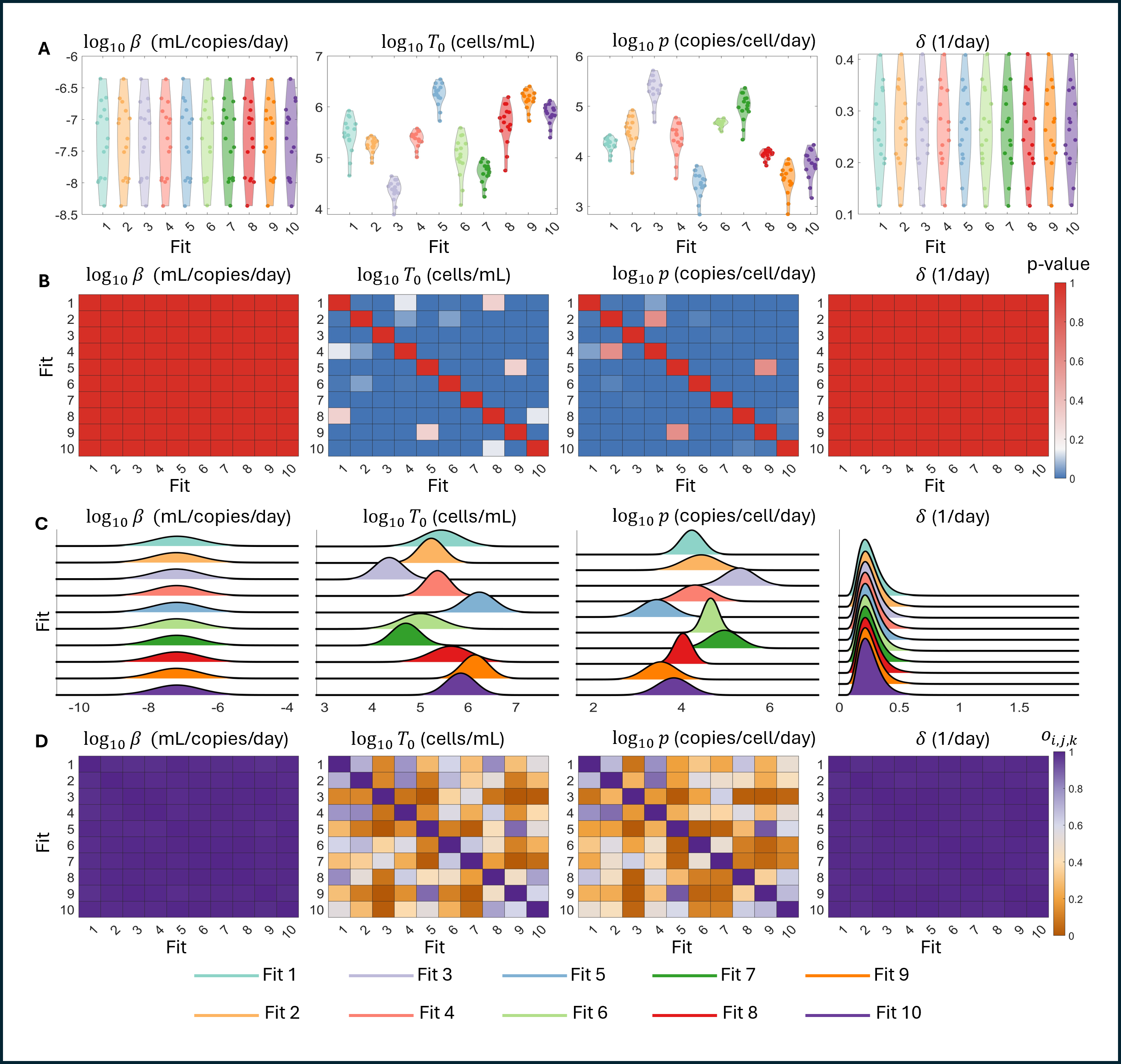}
 \caption{ \textbf{Assessing practical identifiability of the standard viral dynamics model.}  Panel A shows the distribution of the 15 individual estimates for the model parameters $\log_{10} (\beta), \log_{10} (T_0) , \log_{10} (p)$ and $\delta $ for each of the 10 best population fits as dots within each violin plot. Panel B shows the $p$-value corresponding to the pairwise Kolmogorov-Smirnov test between all 10 best fits for each of the four model parameters. Here, $p$-values below the significance threshold of $0.05$ are plotted as blue squares. Panel C shows the 10 estimated population distributions for each of the four model parameters. Panel D shows the overlapping index between each of the pairwise comparisons of the 10 best fits. Here, for each of the $k = 1,2,3,4$ model parameters, pairs of fits $(i,j)$  with an overlapping index $ o_{i,j,k} > 0.5$ are represented by the shades of violet, while pairs with $o_{i,j,k} < 0.5$ are represented by shades of orange. The 10 best fits are distinguished by their colors for the distributions shown in Panels A and C.  }
\label{Fig:HIVIdentifiability}  
\end{figure}

We next analyze the parameter estimates from the 10 best fits. Here, we do not show the distribution of  $V(0),$ as this parameters is directly informed by the synthetic data. In Figure~\ref{Fig:HIVIdentifiability}, we show the individual estimates and the population distributions for the model parameters $\log_{10}(\beta), \ \log_{10}(T_0), \ \log_{10}(p),$ and $\delta$. The distributions of individual estimates for the parameters $\log_{10} \beta$ and $\delta$ are indistinguishable across the 10 best-fits. Accordingly, the Kolmogorov-Smirnov two-sample test does not indicate that there is a significant difference between the individual estimates for these two model parameters. Conversely, the distributions of individual parameter estimates for $\log_{10} T_0$ and $\log_{10} p $ are observably distinct in Figure~\ref{Fig:HIVIdentifiability}A. The Kolmogorov-Smirnov test identifies significant differences between almost all pairs of individual estimates for these two parameters. In Figure~\ref{Fig:HIVIdentifiability}B, the large number of blue squares correspond to pairs of individual estimates with $p < 0.05$. Consequently, our analysis of the individual estimates suggests that $\log_{10} T_0$ and $\log_{10} p $ may not be identifiable at the population level, as these statistically different distributions result in identical viral load dynamics. This conclusion is consistent with earlier structural and practical identifiability analysis of the standard viral dynamics model \citep{Miao2011,Wu2008} which showed that $p$ cannot be identified from viral load data alone. 

We next consider the population distributions, which reflect the conclusions from the individual estimates. Once again, the probability densities of population distributions of $\log_{10} \beta$ and $\delta$ are indistinguishable in Figure~\ref{Fig:HIVIdentifiability}C. As expected from the probability densities for the population distribution of these parameters shown in Figure~\ref{Fig:HIVIdentifiability}C, the overlapping indices for the 10 best fits of $\log_{10} \beta$ and $\delta$ are all greater than 0.97 and 0.98, respectively (Figure~\ref{Fig:HIVIdentifiability}D). Thus, the population distributions for these two parameters have considerable overlap with a correspondingly small total variation distance. Conversely, there is noticeable heterogeneity in the mean and shape of the population distributions of $\log_{10} T_0$ and $\log_{10} p $. This heterogeneity is reflected in the overlapping indices of the population distributions of these two parameters, where the minimum overlap between two distributions is roughly $10^{-3}$ in both cases, which indicates a large total variation distance between the population parameter distributions. Thus, our analysis demonstrates that distinct population distributions for $\log_{10} T_0$ and $\log_{10} p $ can result in nearly identical model predictions and fits to data. This analysis indicates that $\log \beta$ and $\delta$ are identifiable at the population level, while $\log_{10}T_0$ and $\log_{10} p $ are unidentifiable.

\subsubsection{Identifiability of the standard viral dynamics model with a correlation structure}

Further analysis of the identifiability results for the standard viral dynamics model suggest a correlation between $\log T_0$ and $\log p $. This result is unsurprising from the biological interpretation of these parameters, as $\log T_0$ determines the cells available to be infected, while $\log p$ determines the viral production rate. For a given set-point viral load, corresponding to the equilibrium virus concentration observed at later time points, there is a clear relationship between these two mechanisms. For example, fewer infected cells, with a corresponding lower value of $\log T_0$, and a higher production rate $\log p$ can produce the same total amount of virus as more infected cells with a lower production rate, as discussed by \citet{Wu2008}. Further, \citet{Lavielle2016} has shown that some otherwise unidentifiable parameters may be identifiable within a hierarchical framework with an appropriate correlation structure. We therefore allowed for a correlation between these two parameters and once again fit Eq.~\eqref{Eq:TIVDynamicsModel} to the synthetic viral load from $N = 15 $ individuals. As before, we considered the 10 best fits to this synthetic data, with the corresponding population log-likelihoods given in Table~\ref{Table:HIVGoodnessOfFitResultsCorrelation}. We once again note that these 10 bests fit the synthetic data equally well, as measured by a negligible difference in the population $-2\mathcal{LL}$ values, and all 10 fits capture the individual dynamics.

\begin{table}[!ht]
\begin{tabular}{ll|ll|ll|ll|ll}  
Fit  & $-2\mathcal{LL}$  & Fit  & $-2\mathcal{LL}$ & Fit  & $-2\mathcal{LL}$  & Fit  & $-2\mathcal{LL}$  & Fit   & $-2\mathcal{LL}$  \\  \hline
Fit 1 & -217.8  & Fit 2 & -217.8 & Fit 3 & -217.7  & Fit 4 & -217.7 & Fit 5 & -217.7 \\ \hline
 Fit 6 & -217.6  & Fit 7 & -217.6 & Fit 8 & -217.6  & Fit 9 & -217.6 & Fit 10 & -217.6  \\ \hline
\end{tabular}
\caption{ \textbf{Summary of the 10 best fits to the synthetic viral load data}. The 10 best fits to the synthetic viral load data using Eq.~\eqref{Eq:TIVDynamicsModel} and allowing for correlation between $\log T_0$ and $\log p$, as measured by the log-likelihood. Here,  lower values of  $-2\mathcal{LL}$  correspond to a better fit.  }
\label{Table:HIVGoodnessOfFitResultsCorrelation}
\end{table}

\begin{figure}[!h]
\centering
\includegraphics[trim= 4 5 5 10,clip,width=1\textwidth]{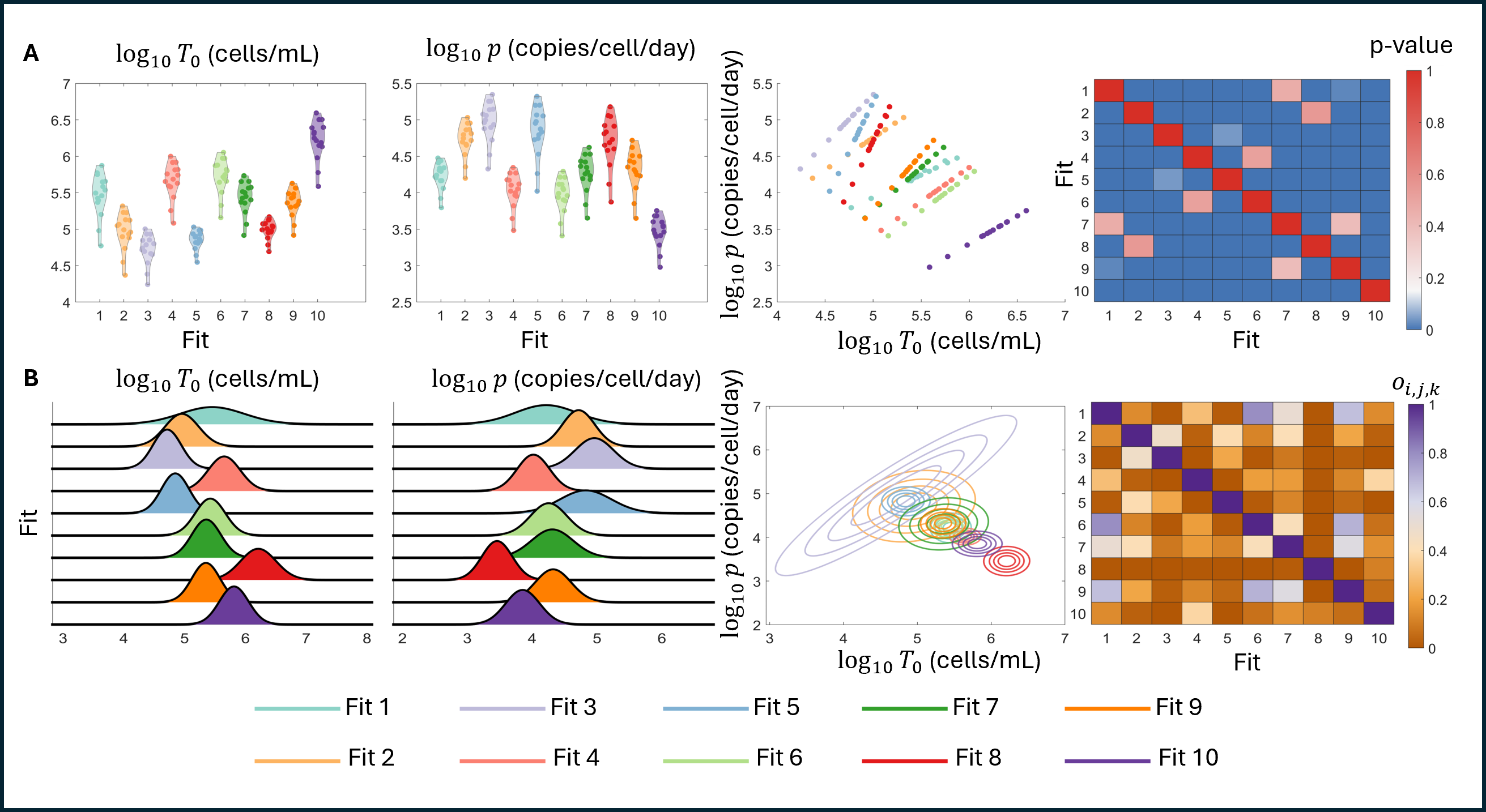}
 \caption{ \textbf{Practical identifiability of the standard viral dynamics model with a correlation between $\log p$ and $\log T_0$}  Panel A shows the distribution of the 15 individual estimates for the model parameters $\log_{10} (\beta), \log_{10} (T_0) , \log_{10} (p)$ and $\delta $ for each of the 10 best population fits as dots within each violin plot. Panel B shows the $p$-value corresponding to the pairwise Kolmogorov-Smirnov test between all 10 best fits for each of the four model parameters. Here, $p$-values below the significance threshold of $0.05$ are plotted as blue squares. Panel C shows the 10 estimated population distributions for each of the four model parameters. Panel D shows the overlapping index between each of the pairwise comparisons of the 10 best fits. Here, for each of the $k = 1,2,3,4$ model parameters, pairs of fits $(i,j)$  with an overlapping index $ o_{i,j,k} > 0.5$ are represented by the shades of violet, while pairs with $o_{i,j,k} < 0.5$ are represented by shades of orange. The 10 best fits are distinguished by their colors for the distributions shown in Panels A and C.  }
\label{Fig:HIVIdentifiabilityCorrelation}  
\end{figure}

We analyse the parameter estimates from these ten fits. As the analysis of the individual and population distributions for $\log(\beta)$ and $\delta$ is unchanged from Figure~\ref{Fig:HIVIdentifiability}, we do not include it here. Rather, we focus on the identifiability of the correlated parameters $\log_{10} T_0$ and $\log_{10} p$. In Figure~\ref{Fig:HIVIdentifiabilityCorrelation}A, we show the individual estimates for these parameters for each of the 10 best fits. Then, for each of the 15 individuals, we plot the individual estimates for $\log_{10} T_0$ against $\log_{10} p$ and note the clear positive correlation for the 15 individuals in each fit, corresponding to a postive correlation between the individual random effects, but an overall negative correlation when considering each of the 10 fits. We further note that the estimates for each fit do not significantly overlap in $\log_{10} T_0 \times \log_{10} p$ space. Accordingly, the MMD two-dimensional test identifies significant differences between nearly all pairs of two-dimensional estimates of $(\log_{10} T_0,\log_{10} p)$, with the blue squares in Figure~\ref{Fig:HIVIdentifiabilityCorrelation}A representing pairs of estimates with $p < 0.05$. This analysis suggests that these two parameters may not be identifiable at the population level, even when allowing for a correlation between them. 

We next consider the population distributions. Once again, we only show the distributions for $\log_{10} T_0$ and $\log_{10} p$ in Figure~\ref{Fig:HIVIdentifiabilityCorrelation}B. Here, we observe the univariate parameter distribtuions reflect the negative correlation structure, with high estimates for $\log_{10} p $ corresponding to lower estimates of $\log_{10} T_0,$ and vice versa. The lack of overlap between the bivariate parameter distributions is reflected in the relatively low overlapping index, and thus large total variation distance, calculated between the population distribution for each pair of fits. Consequently, our analysis at the individual and population level indicates that these two parameters are unidentifiable. 

\section{Discussion}

Nonlinear mixed-effects models are extensively used within the pharmacokinetics modelling community \citep{Traynard2020,Lavielle2014,Clarke1999} and hierarchical models are increasingly common throughout mathematical biology \citep{Ke2022,Gilmore2013,Carpenter2017}. Although this modelling framework benefits from simultaneously considering data from multiple individual participants when estimating parameters, existing approaches to study the identifiability of model parameters relies on a parametric interpretation of the population parameter distributions. Here, we have presented an alternative approach to studying the practical identifiability of a nonlinear mixed-effects model through a nonparametric approach. As we do not consider the parametric representation of the population parameter distributions, our approach is immediately applicable in the case where modellers consider different parameter distributions, or in the fully Bayesian case. Specifically, rather than studying when the parametric representation of two distributions are precisely equal, we focus on determining when two distributions are statistically different. 

We utilize existing statistical techniques to determine whether two distributions differ. At the individual level, we utilize the Kolmogorov-Smirnov two-sample test to test if the individual samples from the underlying population distribution, corresponding to the individuals in the trial, are sufficient to distinguish the underlying population-level distributions. However, this individual-level approach may not distinguish between population distributions that differ at the tails, although other statistical tests, such as the Anderson-Darling or Cram\'{e}r-Von Mises tests, could be used. Consequently, at the population level, we consider the overlapping index to quantify how the probability density functions of two population distributions differ. This approach calculates the total variation distance between the two corresponding probability measures but does not permit a simple test statistic to determine if the two distributions are significantly different, although it would be straightforward to fix a significance threshold in practice. However, our approach to practical identifiability can be easily extended to other statistical frameworks to distinguish between population parameter distributions, including other nonparametric tests that may be more appropriate or informative for a given model.

We illustrated our approach by analysing the identifiability of two important models in pharmacology and viral dynamics. We showed that the Friberg model of chemotherapy-induced neutropenia is identifiable within a hierarchical framework \textit{without} imposing that $k_{prol} = k_{tr}.$ Thus, our work demonstrates that considering multiple individuals allows for the Friberg model to be parameterized without assuming that $k_{tr} = k_{prol}$ \citep{Evans2018,deSouza2017}. Conversely, our results indicate that the standard viral dynamics model remains unidentifiable, even when considering multiple individuals. This population level analysis is consistent with earlier work that considered the identifiability of the standard viral dynamics model when fitting viral load data from a single individual \citep{Wu2008,Cardozo-Ojeda2021}. Consequently, the standard viral dynamics model is an example of where considering multiple individuals does not permit the identification of additional model parameters. Further, our analysis of the standard viral dynamics model indicates that a multi-start fitting approach, as suggested by \citet{Duchesne2021}, may not be sufficient to determine a unique global minimum, as we found over 10 local minima of the log-likelihood that have nearly identical goodness-of-fits. These examples are intended to illustrate the proposed nonparametric framework for assessing practical identifiability. Consequently, we focus on evaluating identifiability given the available data, rather than investigating how trial design factors, such as sample size or sampling frequency, might influence identifiability, although these are important questions for future work.

We have proposed a nonparametric approach to studying practical identifiability within the framework of nonlinear mixed-effects models. While our proposed approach is simple, it has some limitations. For example, our proposed method is computationally demanding as we utilize a multi-start approach to seek the global optima of the log-likelihood. Further complicating this multi-start approach, the parameter estimation problem may be high-dimensional and the objective function is not necessarily convex. Consequently, it is not \textit{a priori} apparent how many runs are necessary to find the global optimum for a given model. Furthermore, we focused on characterizing the population distribution of individual parameters, rather than multivariate joint distributions that would arise from an underlying covariate structure, although we illustrated how our approach extends to the multivariate case in the second example. Nevertheless, there are several areas for future work from both theoretical and practical perspectives. For example, we focused on the identifiability of model parameters for a fixed error model. However, it is possible that using a different error model, which could result in the existence of different local optima, may influence the conclusions of this identifiability analysis \citep{Porthiyas2024,Liu2024,Murphy2024}. Further, this proposed approach to practical identifiability could be implemented within the convergence assessment framework of existing software packages \citep{Monolix2020,Carpenter2017} or the profile likelihood approach to parameter identifiability \citep{Raue2009}.  
 
Despite these limitations, we have developed a simple, nonparametric approach to assess the practical identifiability of hierarchical models. While our examples focused on NLME models, the proposed framework immediately extends to purely Bayesian perspectives \citep{Carpenter2017,Browning2024,Porthiyas2024,Linden2022} and is simple to implement in most programming languages. The framework presented here can be used to assess identifiability in hierarchical models in biology and to identify when a model which is practically non-identifiable at the individual level can become identifiable at the population level.   

\section*{Acknowledgments}
All authors are grateful to the MATRIX mathematical research institute for hosting a one-week residential workshop entitled “Parameter identifiability in mathematical biology” (September 2024) where initial work on this project took place. TC is grateful for partial support from the QJMAM Fund for Applied Mathematics and the London Mathematical Society. 
SH was funded by Wenner-Gren Stiftelserna/the Wenner-Gren Foundations (WGF2022-0044), the Swedish Research Council (project 2024-05621), and the Kjell och M{\"a}rta Beijer Foundation.

\noindent The code underlying these results is available at
\sloppy
\texttt{https://github.com/ttcassid/hierarchical-identifiability}

\section*{Author contributions}
 
\textbf{Conceptualization:} TC. 
\textbf{Formal analysis:} TC.
\textbf{Validation}: TC.
\textbf{Investigation:} TC, MP, STJ, IB, SH.
\textbf{Methodology:} TC, RJM, SH.
\textbf{Software:} TC, MP, STJ.
\textbf{Visualization:} TC.
\textbf{Writing – original draft:} TC.
\textbf{Writing – review \& editing:} All authors.
\textbf{Project administration}: TC.


\section*{Supplemental information}

Here, we use a classical example to demonstrate that a model that is non-identifiable for parameters at the individual level may be identifiable at the population level.   
\subsection*{Identifiability of an exponential growth model}
We consider a phenomenological model of exponential growth where the population size $x_i(t)$ satisfies
\begin{align}
     \TimeDeriv x_i(t) = (a_i+b_i)x_i(t).
     \label{Eq:expgrowth}
\end{align}
 The rates $a_i$ and $b_i$ could, for example, represent cell reproduction and  transwell migration/invasion in \textit{in vitro} growth assays with experimental replicates labeled by indices $i$. Eq.~\eqref{Eq:expgrowth} is a classic example of non-identifiability at the individual level \citep{Lavielle2016}, as it is straightforward to see that the sum of $a_i$ and $b_i$ determines the growth rate of $x_i$. Consequently, there are infinitely many pairs $(a_i,b_i)$ that will lead to identical growth dynamics on the level of individual replicates. However, \citet{Lavielle2016} showed analytically that if the individual-level parameters $a_i$ and $b_i$ are both exponentially distributed with means $\mu_a$ and $\mu_b$, respectively, then the population-level parameters $\mu_a$ and $\mu_b$ can be identified from sufficient data  \citep{Lavielle2016}. Consequently, this simple exponential growth model becomes identifiable at the \textit{population} level under assumptions on the underlying population distribution of $a$ and $b$. To illustrate this result numerically and investigate the impact of population sizes, we generate synthetic data and perform inference in an attempt to identify the population parameters. 
%


\subsection*{Generating virtual replicates and synthetic data} 
Let $P_a$ and $P_b$ be the population distributions for the model parameters $a_i$ and $b_i$, respectively. We follow \citet{Lavielle2016} and assume that $P_a$ and $P_b$ are exponential distributions with respective means $\mu_a=1$ and $\mu_b = 0.1$. We considered $n$ replicates of this simple exponential growth model, where each replicate is generated by sampling individual growth rates $a_i$ and $b_i$ from $P_a$ and $P_b$, respectively. The dynamics of the $i$th replicate, $x_i$, are then given by
\begin{align*}
    x_i(t;a_i,b_i) = x_0 e^{(a_i+b_i)t}, 
\end{align*}
which is immediately implied by Eq.~\eqref{Eq:expgrowth}. For simplicity, we assumed the initial condition $x_0$ to be fixed, known, and common for all replicates $i$.
 
We consider observations of $x_i(t)$ at equally spaced time points $t_j=0,0.2,0.4,0.6,0.8,1$ and perturbed the logarithm of these observations by independent identically distributed (IID) Gaussian noise $\varepsilon_{ij} \overset{\mathrm{IID}}{\sim} N(0,0.025)$, thus generating synthetic data points
\begin{align*}
    y_{ij} = \log( x_i(t_j) ) + \varepsilon_{ij}.
\end{align*} 
Here, we set $\sigma^2=0.025$ and considered population sizes $n = 5, 50, 100,$ and $200$ to illustrate how the identifiability of the population estimates depends on the amount of data available for calibration.

\subsection*{Parameterization of the simple exponential growth model} 
The exponential growth model in Eq.~\eqref{Eq:expgrowth} has an analytic solution. Therefore, we can explicitly derive the likelihood function for the synthetic data points $y_{ij}$, given $\theta=[\mu_a,\mu_b]$. As each individual replicate $i$ is independent, the likelihood function is 
\begin{align*}
    L(\theta) = \prod_{i=1}^n P(y_{i\cdot} | \theta ),
\end{align*}
where $P(y_{i\cdot} | \theta)$ is the probability of observing the data $y_{ij}$ ($j=1,\ldots,m$), given the population parameters $\theta$. We calculate
\begin{eqnarray}
    P( y_{i\cdot} | \theta) &=& \int P(y_{i\cdot} | a_i, b_i) P(a_i, b_i | \theta) da_i db_i \nonumber \\
    &=& \int \prod_{j=1}^m \frac{1}{\sigma\sqrt{2\pi}} \exp\left(-\frac{(y_{ij}-\log[  x_i(t_j;a_i,b_i) ] )^2}{2\sigma^2} \right) P(a_i, b_i | \theta) \d a_i \d b_i \nonumber \\
    &=& C \int  \exp\left(- \frac{1}{2\sigma^2} \sum_{j=1}^m (y_{ij}-\log[ x_i(t_j;a_i,b_i) ] )^2 \right) P(a_i, b_i | \theta) \d a_i \d b_i,\nonumber
\end{eqnarray}
where the constant $C$ is independent of the log-transformed data, $y_{ij}$, but may depend on $\sigma^2$, which we are not estimating. This integral is difficult to evaluate accurately by numerical quadrature. Consequently, we use Monte Carlo integration to estimate $P(y| \theta)$ for the entirety of the synthetic data $y$ via
\begin{equation} \label{eq:MC_integral}
    P( y | \theta) \approx \frac{C}{n} \sum_{i=1}^n  \exp\left(- \frac{1}{2\sigma^2} \sum_{j=1}^m (y_{ij}-\log[ x_i(t_j;a_i,b_i) ])^2 \right)  .
\end{equation} 

\subsection*{Identifying mean growth rates in a simple exponential growth model}
 
\begin{figure}[h!]
    \begin{center}
        \includegraphics[trim= 4 10 5 10,clip,width=1\textwidth]{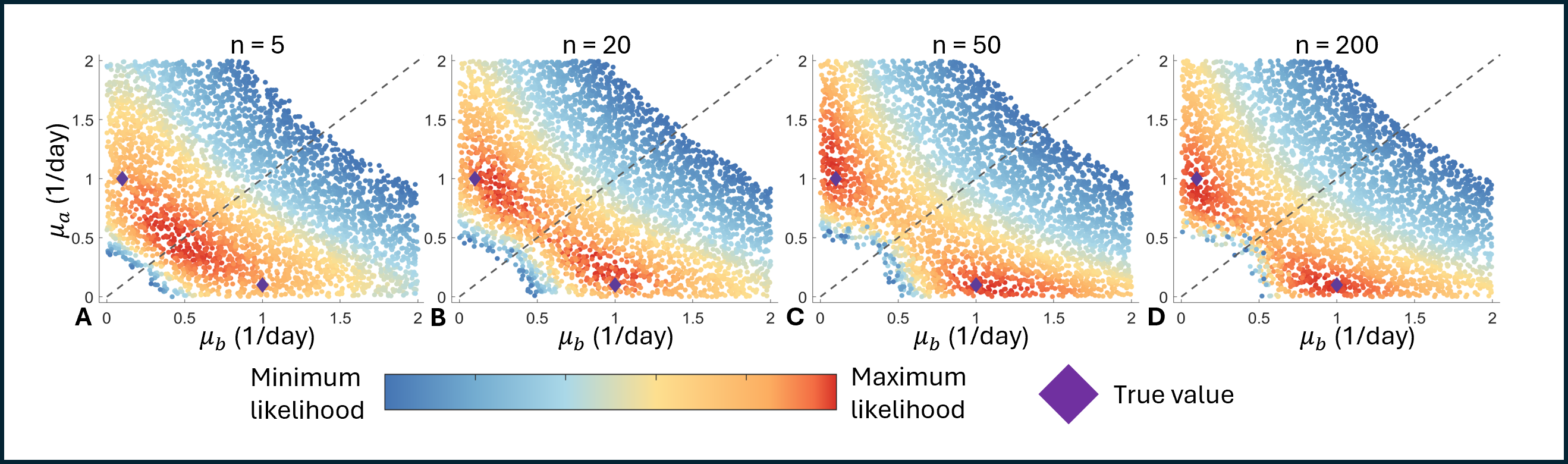}
    \end{center}
    \caption{{\bf The log-likelihood distribution as a function of the mean values of the exponential distributions for the growth rates $a$ and $b$, $\mu_a$ and $\mu_b$.} Panels \textbf{A-D} show the likelihood for $n = 5, 20, 50$, and $200$ virtual replicates. The true value of the mean growth rates, $\mu_a$ and $\mu_b$, are plotted as a diamonds. Since the inference problem is identical after relabelling $a$ and $b$, the pairs $(\mu_a,\mu_b)$ and $(\mu_b,\mu_a)$ both represent true values. The best 4000 samples from the likelihood are shown as filled circles. The likelihood has a reflection symmetry through the line $\mu_a = \mu_b$, shown as a grey dashed line, which corresponds to the relabelling of $a$ and $b$. } 
    \label{Fig:SimpleExponentialGrowth}  
\end{figure}

We compute the log-likelihood in Eq.~\eqref{eq:MC_integral} via Monte-Carlo integration to estimate the population-level parameters.  In Figure~\ref{Fig:SimpleExponentialGrowth}, we plot the log-likelihood function evaluated at 800 uniformly sampled parameter pairs $(\mu_a,\mu_b)$ to estimate the population-level parameters of the simple exponential growth model in Eq.~\eqref{Eq:expgrowth} for population sizes of $n = 5, 20, 50, 200$. 
As the model is clearly unidentifiable at the individual replicate level, it is unsurprising to see that the likelihood does not accurately capture the population-level parameters $\mu_a$ and $\mu_b$ for small population sizes $n = 5$. However, as the population size increases, especially to $n = 50$ and $n = 200$, the likelihood begins to more accurately identify the underlying values of $\mu_a$ and $\mu_b$. 
Figure ~\ref{Fig:SimpleExponentialGrowth} thus  illustrates the analytical results of \citet{Lavielle2016}, and demonstrates that hierarchical parameter estimation can facilitate the identification of model parameters that would otherwise be unidentifiable. 

\section*{Further results on the Friberg Model}
\begin{figure}[!h]
\centering
\includegraphics[trim= 4 5 5 10,clip,width=1\textwidth]{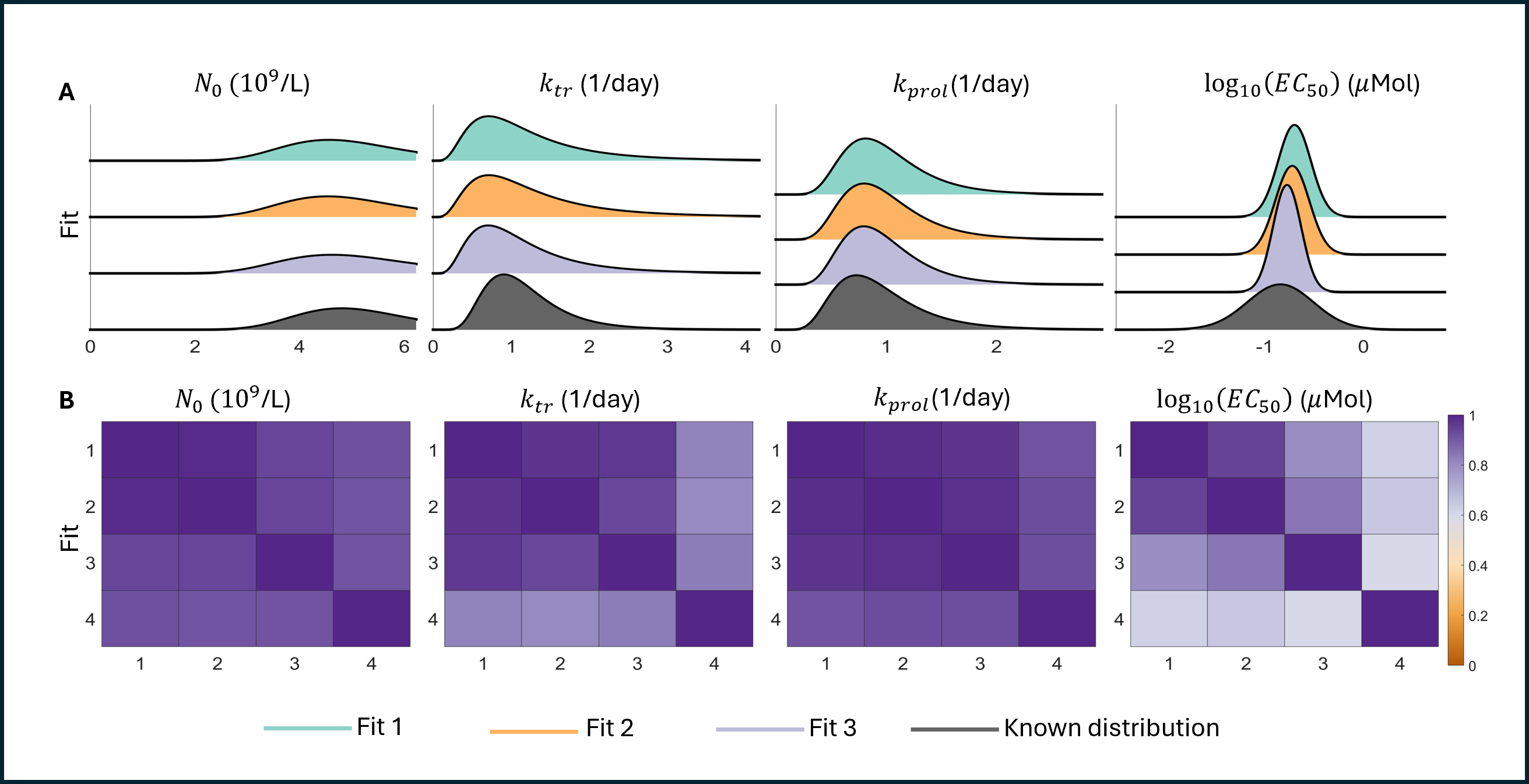}
 \caption{\textbf{Comparison between the estimated and known parameter distributions from the the Friberg model}. Panel A compares the 3 parameter estimates in cluster 1 against the known parameter distributions used to generate the synthetic data for each of the four model parameters. Panel B shows the overlapping index between each of the pairwise comparisons of the 3 parameter estimates in cluster 1 against the known parameter distributions. Here, for each of the $k = 1,2,3,4$ model parameters, pairs of fits $(i,j)$  with an overlapping index $ o_{i,j,k} > 0.5$ are represented by the shades of violet, while pairs with $o_{i,j,k} < 0.5$ are represented by shades of orange. }
\label{Fig:ComparisonFribergIdentifiabilitySI}  
\end{figure}

Here, we present additional results on the identifiability of the Friberg model. We recall that we identified three distinct local minima of the log-likelihood and concluded that fits 1-3 correspond to the global minima of the log-likelihood. In Figure~\ref{Fig:ComparisonFribergIdentifiabilitySI}, we compare the estimated population distributions for $N_0, k_{tr}, k_{prol}$ and $\log_{10}(EC_{50})$ against the known population distributions used to generate the synthetic data. In general, the estimated distributions for fits 1-3 recapture the known underlying distributions. Taken together with our analysis in the main text, we conclude that the Friberg model is identifiable at the population level and the estimated population distributions recapture the underlying true distributions.  

\begin{figure}[!h]
\centering
\includegraphics[trim= 4 5 5 10,clip,width=1\textwidth]{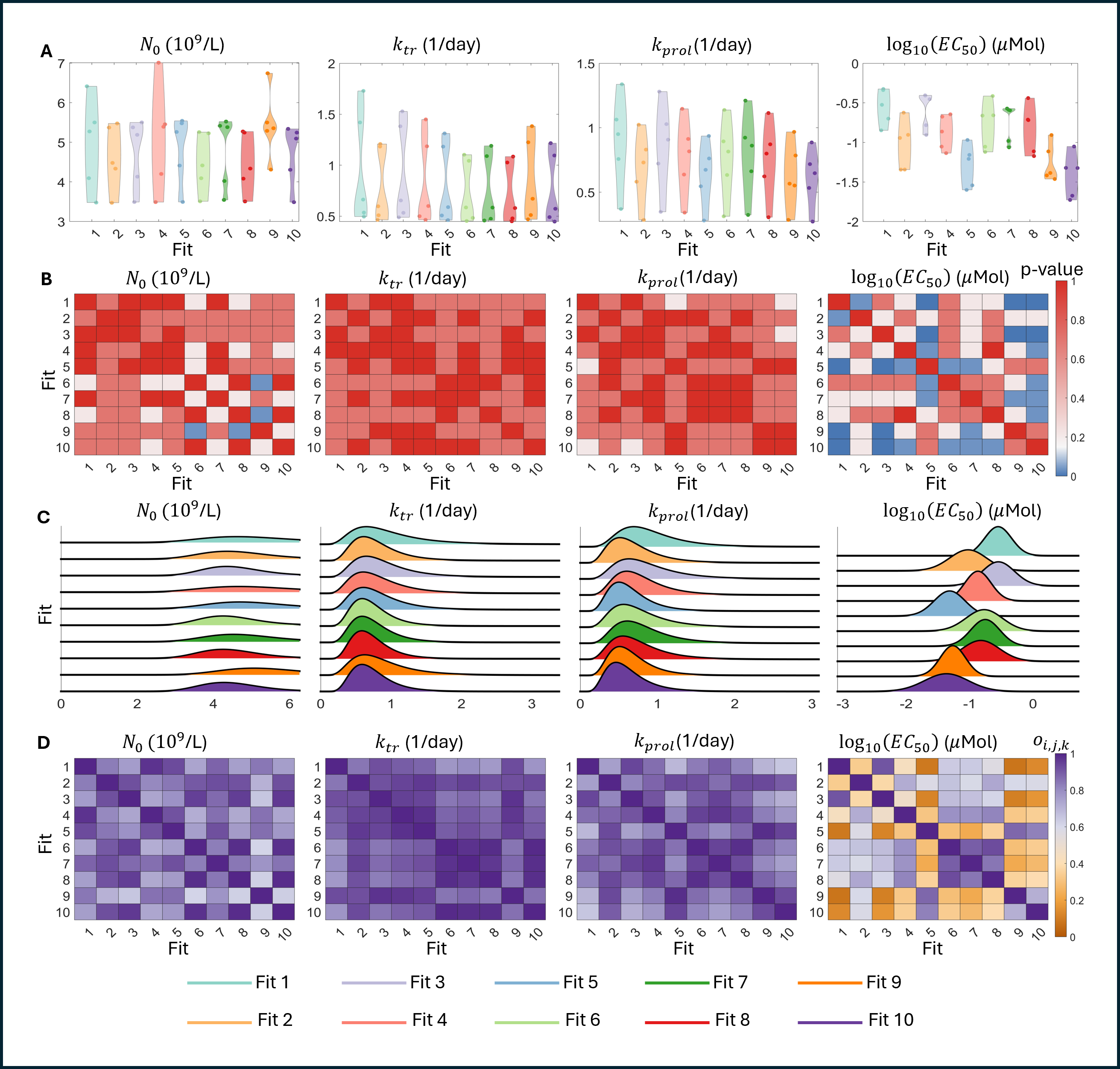}
 \caption{\textbf{Assessing practical identifiability of the Friberg model for $N=5$ virtual participants}.  Panel A shows the distribution of the 15 individual estimates for the model parameters $N_0, k_{tr}, k_{prol}$ and $\log (EC_{50})$ for each of the 10 best population fits as dots within each violin plot. Panel B shows the $p$-value corresponding to the pairwise Kolmogorov-Smirnov test between all 10 best fits for each of the four model parameters. Here, $p$-values below the significance threshold of $0.05$ are plotted as blue squares. Panel C shows the 10 estimated population distributions for each of the four model parameters. Panel D shows the overlapping index between each of the pairwise comparisons of the 10 best fits. Here, for each of the $k = 1,2,3,4$ model parameters, pairs of fits $(i,j)$  with an overlapping index $ o_{i,j,k} > 0.5$ are represented by the shades of violet, while pairs with $o_{i,j,k} < 0.5$ are represented by shades of orange and the three clusters are again highlighted by white boxes. }
\label{Fig:N5FribergIdentifiability}  
\end{figure}

In the previous analysis, we considered $N=15$ participants. To analyse if the Friberg model remains practically identifiable at the population level for a smaller number of participants, we consider $N =5$ virtual participants. We once again considered 100 initial estimates for the model parameters and fit the synthetic data from these initial estimates. As in the main text, we consider the 10 best fits to the data, as measured by AIC (or log-likelihood). As in the main text, the model captured the individual dynamics for all 5 virtual participants. We therefore considered the individual and population estimates for the model parameters. These parameter estimates are shown in Figure~\ref{Fig:N5FribergIdentifiability}. Broadly speaking, the Kolmogorov-Smirnov test does not identify significant differences between the distributions of individual estimates for $N_0,\ k_{tr}$ and $k_{prol}$. However, there are some pairwise significant differences between the distributions of individual estimates for $\log_{10}(EC_{50})$. These results are mirrored in the population distributions, where there are large overlaps between $N_0,\ k_{tr}$ and $k_{prol}$. Conversely, some pairs of population distributions of $\log_{10}(EC_{50})$ do not have large overlap. Consequently, these results suggest that the population distribution of $\log_{10}(EC_{50})$ is poorly identified when fitting the model to $N=5$ participants. 

\end{document}